\documentclass[12pt]{article}
\pdfoutput=1
\usepackage{putex}
\usepackage{amsmath,hyperref,amssymb,amsfonts,bm,amscd,slashed,ytableau,upgreek}
\usepackage{tikz}
\usepackage{cite}
\usepackage{graphicx}
\usepackage{multirow}
\usepackage{verbatim}
\usepackage{appendix}
\usepackage{fancybox}
\usepackage{array}
\usepackage{url}
\usepackage{float}
\usepackage{mathtools}
\usepackage{mathrsfs}
\usepackage{bbm}
\usepackage[bbgreekl]{mathbbol}

\DeclareSymbolFontAlphabet{\mathbbm}{bbold}
\DeclareSymbolFontAlphabet{\mathbb}{AMSb}

\DeclareMathAlphabet{\mathpzc}{OT1}{pzc}{m}{it}

\newcommand{\bea}{\begin{eqnarray}}
\newcommand{\eea}{\end{eqnarray}}
\newcommand{\be}{\begin{equation}}
\newcommand{\ee}{\end{equation}}

\def \beaa {\begin{equation}\begin{aligned}}
\def \eeaa {\end{aligned}\end{equation}}

\newcommand{\Z}{{\mathbb Z}}
\newcommand{\R}{{\mathbb R}}
\newcommand{\C}{{\mathbb C}}
\newcommand{\bP}{{\mathbb P}}

\newcommand{\Q}{{\mathbb Q}}

\newcommand{\cF}{{\mathcal{F}}}
\newcommand{\cN}{{\mathcal{N}}}

\newcommand{\dd}{{\rm d}}

\def\Tr{{\rm Tr \,}}

\def\tilde{\widetilde}
\def\hat{\widehat}
\def\bar{\overline}

\def\cA{{\mathcal A}}
\def\cB{{\mathcal B}}

\def\cE{{\mathcal E}}
\def\cF{{\mathcal F}}
\def\cG{{\mathcal G}}
\def\cH{{\mathcal H}}
\def\cI{{\mathcal I}}
\def\cJ{{\mathcal J}}

\def\cL{{\mathcal L}}
\def\cM{{\mathcal M}}
\def\cN{{\mathcal N}}
\def\cO{{\mathcal O}}
\def\cP{{\mathcal P}}
\def\cQ{{\mathcal Q}}
\def\cR{{\mathcal R}}

\def\cT{{\mathcal T}}

\def\cV{{\mathcal V}}
\def\cW{{\mathcal W}}

\def\bE{{\mathbb E}}
\def\bfT{{\mathbf T}}

\def\sD{{\mathscr D}}

\def\sz{{\mathpzc{Z}}}

\renewcommand{\bar}{\overline}
\renewcommand{\hat}{\widehat}

\numberwithin{equation}{section}
\begin{document}

\institution{SCGP}{Simons Center for Geometry and Physics,\cr Stony Brook University, Stony Brook, NY 11794-3636, USA}
	
\title{Interfaces and Quantum Algebras, II:\\ Cigar Partition Function}
\authors{Mykola Dedushenko\worksat{\SCGP} and Nikita Nekrasov\worksat{\SCGP}}
	
\abstract{The supersymmetric cigar (half-)index or cigar partition function of 3d $\mathcal{N}=2$ gauge theories contains a wealth of information. Physically, it captures the spectrum of BPS states, the non-perturbative corrections to various partition functions, the effective twisted superpotential and the data of supersymmetric vacua. Mathematically, it defines the K-theoretic Vertex counting vortices/quasimaps, and connects to quantum K-theory, as well as elliptic cohomology and stable envelopes. We explore these topics from the physics standpoint, systematically developing the foundations and explaining various mathematical properties using the quantum field theory machinery.}

\date{}
	
\maketitle
	
\tableofcontents
	
\section{Introduction}

Quantum field theory is a rich source of mathematical problems, including formulating its own principles.
It is believed that whatever quantum field theory is, it is probably defined on a variety of geometric backgrounds, both real and complex, contains its own deformation theory, and is decorated with a plethora of algebraic structures. A class of quantum field theories admits finite dimensional approximations, or solvable sectors, which might be described by quantum integrable systems. Also, it is believed that certain quantum field theories can be fused, connected via some sort of interfaces, in which case the integrable subsectors get united, sometimes in a way that a large quantum algebra, such as a Yangian, acts in the irreducible way on the union. 

Although the current paper can be read independently, we should mention that in our previous work \cite{Dedushenko:2021mds}, we started exploring these structures in the class of three dimensional supersymmetric gauge theories, namely, those with ${\cN}=4$ supersymmetry softly broken down to $\cN=2$ by some background fields or twisted masses. In that work, we began by constructing supersymmetric interfaces providing the physical counterpart of stable envelopes \cite{Maulik:2012wi, Aganagic:2016jmx}, and serving as building blocks for more elaborate constructions. Closely related explorations were also reported in \cite{Bullimore:2021rnr}.

One of the ways supersymmetric quantum field theory connects to conventional mathematics is through its vacuum structure. The vacua of supersymmetric theory can be related to some (extraordinary) cohomology theory. For example, supersymmetric quantum mechanics of a particle moving on a Riemannian manifold $M$ is connected to harmonic $L^2$ spinors on $M$; extended supersymmetry produces de Rham cohomology. Going up in dimension of spacetime brings about $K$-theory and elliptic cohomology. 
Cohomology theory usually comes with some ring structure, which in physics realization originates in the state-operator correspondence, and the Euclidean time evolution projecting onto the harmonic ground states. 
Non-perturbative physics deforms the classical ring structure. For example, two dimensional sigma model instantons are responsible for quantum cohomology rings. The lifts of two dimensional theory to three and four dimensional QFT compactified on a circle and a torus, respectively,  are expected \cite{Baulieu:1997nj, Jockers:2018sfl} to produce the quantum $K$-theory and (conjectural) quantum elliptic cohomology. 

Quantum field theory obeying Poincare invariance in flat space admits different Hamiltonian descriptions when formulated on a curved spacetime. In this way, what looks like a transition between two vacua $a$ and $b$ gets mapped to a trace on a Hilbert space of $ab$-solitons. More generally, the plethora of boundary conditions on spacetimes of dimension two and higher (and in different co-dimension) suggests the existence of higher multiplications on the spaces of supersymmetric states, making them into $\infty$-algebras, cf. \cite{Gaiotto:2015aoa}.

Correlation functions of supersymmetric operators often have limited or no dependence on geometry, such as the positions of local operators, shapes of line defects, etc. Depending on the type of supercharge the operators commute with, the correlation functions may be holomorphic in the operator positions or, more generally, holomorphic in the complexified geometry \cite{Witten:1986bf} or in some external parameters, such as coupling constants, equivariant flavor or K{\"a}hler moduli \cite{Beem:2013sza, Dedushenko:2017tdw}. 

Sometimes the same holomorphic functions emerge from two different realizations, e.g. as conformal blocks of some two-dimensional conformal field theory, and as instanton partition functions of some four dimensional supersymmetric gauge theory.

Quite general class of such correlation functions is provided by $J$-functions \cite{Givental:1996} of two dimensional supersymmetric sigma models. They obey a system of differential equations (also known as the $t$-part of the $tt^*$-structure \cite{Cecotti:1991me}), closely resembling Knizhnik-Zamolodchikov-Bernard equations \cite{Knizhnik:1984nr, Bernard:1987df}
of two dimensional conformal field theory with some current algebra, or Belavin-Polyakov-Zamolodchikov equations \cite{Belavin:1984vu} of two dimensional conformal field theory  \cite{Braverman:2010ei, Nekrasov:2015wsu, Nekrasov:2017gzb, Nekrasov:2021tik}. The remarkable feature of both sets of equations is their $\lambda$-connection form, i.e. they form a pencil of flat connections \cite{Losev:2000qq} over some base, whose connection form, in a special coordinate system, depends linearly on auxiliary parameter \cite{Dubrovin:1992dz}. In the 2d CFT realization this parameter is related to the level of current algebra, it measures how quantum the system is. 
In the sigma model realization this parameter is equivariant parameter corresponding to the rotational symmetry of the worldsheet. That quantization parameter of one theory maps to the roational equivariant parameter of another is unexpected at first. However, if one interprets quantization as the path integration over the space of loops of some
classical configuration space, such a connection becomes natural.  

In the present work, we discuss the physical setup in which $J$-functions appear in two and three dimensional supersymmetric gauge theories, focusing on the three-dimensional case. We find it useful to subject our theory to a specific curved background, which is promoted to the supergravity background in which the theory possesses conserved supercharge \cite{Closset:2012ru} allowing the exact evaluation of path integral. 
By exploiting the interpolating features of this background, we find interesting connections between the elliptic cohomology and quantum $K$-theory of Nakajima varieties, stable envelopes of \cite{Maulik:2012wi, Aganagic:2016jmx}, and clarify certain subtle points along the way. 

Our methods are more general, so in principle we can also study the theories which represent gauge theories in higher space-time dimensions. For example, using the ADHM construction one can encode the Witten index of a
$6+1$ or $8+1$ dimensional gauge theory in a kind of cigar partition function we study below. The algebraic structures exhibited by this setup are believed to be identical to those of a $2+1$ dimensional theory. What knows about the space-time dimension is the type of representation of the quantum algebra. 

In Section \ref{sec:cigarBG} we begin by constructing the supersymmetric background in question. It is given by a squashed version of the superconformal (half-)index in three dimensions \cite{Gadde:2013wq,Dimofte:2017tpi}. It can be viewed as a three-dimensional uplift of the squashed two-sphere of \cite{Gomis:2012wy}. In \emph{loc.cit.} its partition function was independent of the squashing parameter, providing an interpolation between the round sphere and the $tt^*$ setup in conformal limit. Our three-dimensional version has the same property. In particular, considering the squashed index $S^1\times S^2_b$ or the squashed half-index background $S^1\times HS^2_b$, where $HS^2_b$ is a squashed hemisphere (half-ellipsoid), we observe an interesting interpolation property. The supercharge preserved by this background looks like the 3d A-model supercharge $\cQ_A$ \cite{Closset:2017zgf,Benini:2015noa} close to the pole of $HS^2$ (which we also call the North pole, or the tip), while near the equator it becomes the supercharge $\cQ$ used in \cite{Dedushenko:2021mds} to study the elliptic stable envelopes. The cohomology of $\cQ_A$ is where one finds the quantum K-theory of the Higgs branch $X$ in 3d, whereas the cohomology of $\cQ$ contains classical elliptic cohomology classes of $X$. This interpolation property is the key to many of our observations.

In Section \ref{sec:vertexF} we study the Vertex function introduced in \cite{Okounkov:2015spn}. The main outcome of this section is that the Vertex $V$ is computed by the topologically twisted index on $S^1\times_\varepsilon \C\bP^1$ (in the sense of 3d A-twist). We prove it using the supersymmetric localization and the topological invariance of twisted index. At the same time, the untwisted half-index of Section \ref{sec:cigarBG} computes a closely related quantity $\tilde{V}$ that differs from $V$ by normalization. This minor difference is quite significant: While $V$ is valued in the equivariant K-theory of the Higgs branch $X$, $\tilde{V}$ is most naturally viewed as the elliptic cohomology class, possibly decorated by a descendant sitting at the pole of $HS^2_b$.

In Section \ref{sec:wallcross} we study the behavior of $\tilde{V}$ under the mirror symmetry and symplectic duality. We show that the elliptic stable envelopes, via their physical realization in \cite{Dedushenko:2021mds}, describe the transformation of $\tilde{V}$ under the symplectic duality, or physically, Higgs/Coulomb phase transition.

Finally, in Section \ref{sec:observables} we study the supersymmetric surface and line operators admitted by $S^1\times HS^2_b$. The surface operators act on the elliptic cohomology class produced by $\tilde{V}$, while the line operators represent K-theory classes that can decorate $\tilde{V}$ at the pole of $HS^2_b$. We prove some properties of these observables, and connect lines to quantum K-theory. In the Outlook section, we mention a couple of open problems.

{\bf Acknowledgements} We are grateful to C.~Closset, K.~Hori, Z.~Komargodski, A.~Okounkov, A.~Smirnov, D.~Zhang for discussions.

{\bf Note:} Submission of this work is being coordinated with \cite{ToCoordinate}.

\section{3D cigar background}\label{sec:cigarBG}
The main object that we study in this paper is the index counting holomorphic quasi-maps \cite{Bradlow,Losev:1999nt,CIOCANFONTANINE201417,Okounkov:2015spn} from $\C \mathbb{P}^1$ into the Higgs or Coulomb branch of vacua. It is known under various names in the literature: the vortex partition function \cite{Shadchin:2006yz,Dimofte:2010tz,Bonelli:2011fq,Bonelli:2013mma}, the holomorphic block \cite{Beem:2012mb}, the homological block \cite{Gukov:2017kmk} (in a narrower context), the half-index \cite{Gadde:2013wq}, or the Vertex function \cite{Okounkov:2015spn}. Not all of these objects are precisely equal, but they are very closely related and, in most cases, proportional to each other. Physically, they all involve counting BPS states of a 3D $\cN=2$ on some sort of cigar geometry, and hence are captured by the appropriate partition function on $S^1 \times_\varepsilon C$. Here $C$ is the ``cigar'', $\varepsilon$ means that as we go around the $S^1$, the cigar is rotated by $2\pi\varepsilon$, and this space is sometimes called ``Melvin cigar'' \cite{Cecotti:2010fi}. In the cases of vortex partition functions, holomorphic blocks, homological blocks, and Vertex functions, $C$ carries the ``3D A-twist'' (see \cite{Closset:2017zgf} on this twist). In the case of half-index, $C$ is the round hemisphere without a twist, which we may call the conformal background, since it is the same as one half of the $S^1\times_\varepsilon S^2$ background that captures the superconformal index.

We add to this list yet another object, -- a squashed half-index, i.e., deformed version of the conformal half-index background. It can also be viewed as the squashed index (properly introduced below as well,) cut in half. It coincides with the usual half-index, which we prove by showing that squashing is a trivial deformation of the transversally holomorphic foliation (THF) determining the SUSY background \cite{Closset:2012ru,Closset:2013vra}. However, the squashing deformation allows (in analogy with the 2D $tt^*$ case of \cite{Gomis:2012wy}) to connect it with other quantities on the list. In later sections, we will use Janus interfaces to transform the half-indices across phase boundaries, specifically between the Higgs and Coulomb phases, thus giving physics interpretation to some of the mathematical results from \cite{Aganagic:2016jmx,Aganagic:2017smx}.
 
\subsection{Squashed 3D $\cN=2$ index}\label{sec:squashedIndex}
Let us start with the squashed index in 3d $\cN=2$ theories. The Gomis-Lee background \cite{Gomis:2012wy} for 2d $\cN=(2,2)$ theories formulated on the squashed two-sphere $S^2_b$ has a natural lift to three dimensions, giving the squashed index geometry\footnote{A similar background on $S^1\times\R\mathbb{P}^2_b$ was written in \cite{Tanaka:2014oda}.} $S^1\times S^2_b$. The $S^1$-invariant squashed two-sphere metric is written following \cite{Gomis:2012wy} as
\begin{align}
\dd s_b^2 = f(\theta)^2 \dd\theta^2 + \ell^2 \sin^2\theta \dd\varphi^2,
\end{align}
where $f(\theta)$ is an arbitrary smooth function obeying $f(0)=f(\pi)=\ell$, so that there are no conical singularities at the poles. A useful model to keep in mind is the one from \cite{Gomis:2012wy}:
\begin{equation}
\label{f_of_theta}
f(\theta)^2 = \tilde{\ell}^2 \sin^2\theta + \ell^2\cos^2\theta,
\end{equation}
where $\ell$ and $\tilde{\ell}$ are the two parameters of the ellipsoid (or rather spheroid):
\begin{equation}
\frac{x_1^2 + x_2^2}{\ell^2} + \frac{x_3^2}{\tilde{\ell}^2} = 1,
\end{equation}
where the Cartesian coordinates are related to the angular variables on the spheroid via
\begin{equation}
x_1=\ell\sin\theta\cos\varphi,\quad x_2 = \ell\sin\theta\sin\varphi,\quad x_3=\tilde{\ell}\cos\theta,
\end{equation}
and the ``squashing parameter'' is
\begin{equation}
b = \frac{\ell}{\tilde{\ell}}.
\end{equation}
On the three-manifold $S^1\times S^2_b$, we could take the product metric. We will be slightly more general, and consider the fiber product $S^1\times_\varepsilon S^2_b$, where the spheroid is rotated by $2\pi \varepsilon$ as we go around the $S^1$:
\begin{equation}
\dd s^2_3 = f(\theta)^2 \dd\theta^2 + \ell^2 \sin^2\theta (\dd\varphi + \varepsilon \dd\alpha)^2 + \beta^2\dd\alpha^2,
\end{equation}
where $\alpha\in [0,2\pi]$ is the angular coordinate on $S^1$. The local frame fields in the chart that excludes the poles of $S^2_b$ can be chosen as:
\begin{equation}
e^{\hat 1} = f(\theta)\dd\theta,\quad e^{\hat 2} = \ell \sin\theta (\dd\varphi + \varepsilon\dd\alpha),\quad e^{\hat 3}=\beta\dd\alpha.
\end{equation}
The corresponding spin connection is
\begin{equation}
\omega \equiv \omega_{\hat{1}\hat{2}} = -\frac{\ell\cos\theta}{f(\theta)}(\dd\varphi + \varepsilon\dd\alpha).
\end{equation}
Coupling to the curved background involves choosing a $U(1)_R$ R-symmetry, and turning the following background for its gauge field $A^{(R)}$:
\begin{equation}
\label{Rbg}
A^{(R)} = \frac12 \left(1-\frac{\ell}{f(\theta)} \right)\dd\varphi + \frac{i\beta-\varepsilon\ell}{2 f(\theta)}\dd\alpha.
\end{equation}
The generalized Killing spinor equations on this background are
\begin{equation}
D_\mu\epsilon = \frac1{2f(\theta)}\gamma_\mu \gamma^{\hat 3}\epsilon,\quad D_\mu\bar\epsilon = -\frac1{2f(\theta)}\gamma_\mu \gamma^{\hat 3}\bar\epsilon,
\end{equation}
where $\gamma^{\hat 3}$ is the gamma matrix in the circle direction, written in the local orthonormal frame, and $\gamma_\mu$ is written in the coordinate frame, i.e., $\gamma_\mu = e_\mu^{\hat a}\gamma_{\hat a}$. In our conventions, $\gamma_{\hat a}$ are given by the Pauli matrices. The covariant derivatives include the spin connection $\nabla_\mu$ and the R-symmetry gauge field:
\begin{equation}
D_\mu\epsilon = (\nabla_\mu - iA^{(R)}_\mu)\epsilon,\quad D_\mu\bar{\epsilon} = (\nabla_\mu + iA^{(R)}_\mu)\bar\epsilon.
\end{equation}
The background preserves two supercharges, parameterized by the following spinors:\footnote{The description that often appears in the literature \cite{Imamura:2011su} concerns theory in the twisted sector for the R-symmetry, in which case the spinors $(\epsilon,\bar\epsilon)$ are not single valued as functions of $\alpha\in S^1$. Such a formulation is related to ours by the R-symmetry gauge transformation.}
\begin{align}
\epsilon &= e^{-\frac{i}{2}\theta\gamma^{\hat 2}} e^{\frac{i}{2}\varphi \gamma^{\hat 3}}\epsilon_o,\quad \text{where} \ \gamma^{\hat 3}\epsilon_o=\epsilon_o,\cr
\bar\epsilon &= e^{\frac{i}{2}\theta\gamma^{\hat 2}} e^{\frac{i}{2}\varphi \gamma^{\hat 3}}\bar\epsilon_o,\quad \text{where} \ \gamma^{\hat 3}\bar\epsilon_o=-\bar\epsilon_o.
\end{align}
Let us denote the supercharges corresponding to $\epsilon_o={1\choose 0}$ and $\bar\epsilon_o={0\choose 1}$ by $\hat{Q}$ and $\hat{S}$, respectively. The anticommutator $\{\hat{Q},\hat{S}\}$ is proportional to $D_\alpha - (\varepsilon + i\frac{\beta}{\ell})D_\varphi + \beta\sigma\cos\theta$, which interpolates between $D_{\bar z}$ at the equator $\theta=\frac{\pi}{2}$ and $D_\alpha + \beta\sigma$ near the tip $\theta=0$, since $D_\varphi$ acts trivially there (here $\sigma$ is a real vector multipelt scalar).

In fact, $Q=\hat{Q}+\hat{S}$ has an interesting property: It interpolates between the 3d A-twist supercharge $\cQ_A$ at the North pole and the supercharge $\cQ$ at the equator, the latter playing central role in the relation to elliptic cohomology and stable envelopes in \cite{Dedushenko:2021mds}. We will comment more on this later, but one can already foresee that this property underlies the main use of this background in connecting the K-theoretic and elliptic observables.

SUSY of the 3d $\cN=2$ vector multiplet receives corrections due to the curvature:
\begin{align}
\delta A_\mu &= \frac{i}{2} (\bar\epsilon \gamma_\mu\lambda + \epsilon\gamma_\mu \bar\lambda),\quad \delta \sigma = \frac12 (\bar\epsilon\lambda - \epsilon\bar\lambda),\cr
\delta\lambda&= -\frac12 F_{\mu\nu} \gamma^{\mu\nu}\epsilon - D\epsilon + i\partial_\mu\sigma \gamma^\mu\epsilon + \frac{i}{f(\theta)}\sigma \gamma_{\hat3} \epsilon,\cr
\delta\bar\lambda&= -\frac12 F_{\mu\nu}\gamma^{\mu\nu}\bar\epsilon + D \bar\epsilon -i\partial_\mu \sigma \gamma^\mu \bar\epsilon + \frac{i}{f(\theta)}\sigma \gamma_{\hat3}\bar\epsilon,\cr
\delta D&= -\frac{i}{2}\bar\epsilon \gamma^\mu D_\mu\lambda + \frac{i}{2} \epsilon \gamma^\mu D_\mu\bar\lambda + \frac{i}{2}\bar\epsilon [\lambda,\sigma] + \frac{i}{2}\epsilon[\bar\lambda,\sigma] + \frac{i}{4 f(\theta)}(\bar\epsilon \gamma_{\hat3} \lambda + \epsilon\gamma_{\hat3}\bar\lambda).\cr
\end{align}
The vector multiplet action also receives a few corrections in our background:
\begin{align}
\cL_{\rm v}&=\Tr\Big[ \frac12 F^{\mu\nu}F_{\mu\nu} + D^\mu\sigma D_\mu\sigma+ D^2 + \frac{\sigma^2}{f(\theta)^2}-\epsilon^{\mu\nu\rho}F_{\mu\nu}D_\rho\sigma - \frac1{f(\theta)}\epsilon^{\mu\nu \hat3}F_{\mu\nu}\sigma + \frac{2}{f(\theta)}\sigma D_{\hat3} \sigma\Big]\cr
&+\Tr\Big[ i\lambda\gamma^\mu D_\mu\bar\lambda + i\lambda[\bar\lambda,\sigma] - \frac{i}{2f(\theta)}\bar\lambda\gamma_{\hat3} \lambda \Big].
\end{align}
The chiral multiplet SUSY and action are:
\begin{align}
\delta \phi &=\bar\epsilon \psi,\quad \delta \bar\phi=\epsilon \bar\psi,\cr
\delta\psi&=i \gamma^\mu \epsilon D_\mu\phi + i\epsilon\sigma\phi + \frac{i\Delta}{f(\theta)} \gamma_{\hat3}\epsilon\phi + \bar{\epsilon} F,\cr
\delta\bar\psi&=i\gamma^\mu \bar\epsilon D_\mu\bar\phi + i\bar\phi \sigma\bar\epsilon - \frac{i\Delta}{f(\theta)}\bar\phi \gamma_{\hat3} \bar\epsilon + \bar{F}\epsilon,\cr
\delta F&= \epsilon(i\gamma^\mu D_\mu\psi -i\sigma\psi -i\lambda\phi) +\frac{i}{2f(\theta)}(2\Delta-1)\epsilon\gamma_{\hat3}\psi,\cr
\delta \bar{F}&=\bar\epsilon(i\gamma^\mu D_\mu\bar\psi-i\bar\psi\sigma+i\bar\phi\bar\lambda)-\frac{i}{2f(\theta)}(2\Delta-1)\bar\epsilon\gamma_{\hat3}\bar\psi,
\end{align}
\begin{align}
\cL_{\rm c}&=-\bar\phi D^\mu D_\mu\phi +\bar\phi \sigma^2 \phi + i\bar\phi D\phi + \bar{F} F + \frac{1-2\Delta}{f(\theta)}\bar\phi D_{\hat3}\phi + \frac{\Delta-\Delta^2}{f(\theta)^2} \bar\phi \phi - \frac{\Delta W_{\theta\varphi} \cos\theta}{\ell\sin\theta f(\theta)}\bar\phi\phi \cr
&-i\bar\psi\gamma^\mu D_\mu \psi + i\bar\psi\sigma\psi + \frac{i(1-2\Delta)}{2f(\theta)}\bar\psi\gamma_{\hat3}\psi +i\bar\psi\lambda\phi -i\bar\phi\bar\lambda\psi,
\end{align}
where $W=\dd A^{(R)}$ is the R-symmetry field strength, and $\Delta$ denotes the R-charge of $\phi$. Both $\cL_{\rm v}$ and $\cL_{\rm c}$ are exact with respect to each of the two preserved supercharges. Finally, the Lagrangian density capturing the superpotential term is the same as in flat space. As for the mass and FI terms, we will consider them in more detail below.

\subsubsection{Squashing $S^2$ is a trivial deformation of the THF}\label{sec:THF}
In the two-dimensional case of \cite{Gomis:2012wy}, the sphere partition function was independent of the squashing parameter $b$. More generally, one would expect independence of $f(\theta)$. This property survives the lift to three dimensions, which we are going to prove now. There is a well-developed machinery for constructing supersymmetric backgrounds in 3d $\cN=2$ \cite{Closset:2012ru,Closset:2013vra}, at least those that fit within the new minimal supergravity, following the philosophy of \cite{Festuccia:2011ws}. Our background falls into such category, and we find that the background supergravity fields $(H, A^{(R)}_\mu, V_\mu)$, in the conventions of \cite{Closset:2013vra}, are: $H=0$, their $A^{(R)}$ coincides with our $A^{(R)}$, and $V=- \frac{i\beta}{f(\theta)}\dd\alpha$.\footnote{In \cite{Closset:2013vra}, there is also $A_\mu$, which is related to the other fields by $A^{(R)}_\mu=A_\mu - \frac{3}{2} V_\mu$.} This indeed corresponds to the deformation of the ordinary index background. To understand the non-trivial (i.e., not Q-exact) deformations, one has to study the corresponding transversally holomorphic foliation (THF). Let us focus on the supercharge parameterized by $\epsilon_o={1\choose 0}$. Construct its associated one-form:
\begin{equation}
\eta_\mu = \frac1{|\epsilon|^2}\epsilon^\dagger \gamma_\mu \epsilon,
\end{equation}
and the corresponding vector field $\xi^\mu=\eta^\mu$. We find:
\begin{align}
\eta&=f(\theta)\sin\theta\, \dd\theta + \beta\cos\theta\, \dd\alpha,\cr
\xi &=\frac{\sin\theta}{f(\theta)} \partial_\theta + \frac{\cos\theta}{\beta} \left(\partial_\alpha - \varepsilon\partial_\varphi \right).
\end{align}
One also defines a tensor $\Phi^\mu{}_\nu=-\varepsilon^\mu{}_{\nu\rho}\xi^\rho$. Then $\xi$ determines a 1d foliation, the THF, with $\Phi^\mu{}_\nu$ playing the role of complex structure in the directions transverse to $\xi$, since it obeys:
\begin{equation}
\Phi^\mu{}_\nu \Phi^\nu{}_\rho = -\delta^\mu{}_\rho +\xi^\mu\eta_\rho.
\end{equation}
We write these data in the adapted coordinates $(t,z,\bar{z})$, which can be chosen as follows:
\begin{align}
t &= \beta\alpha - F(\theta),\quad z=e^{-\frac{\beta}{\ell}\alpha +i(\varphi+\varepsilon\alpha)}G(\theta),\cr
F(\theta)&=\int_{\theta_0}^\theta \frac{f(\theta)(\cos\theta-1)}{\sin\theta}\dd\theta,\quad G(\theta)=e^{\int_{\theta_0}^{\theta} \frac{f(\theta)\cos\theta}{\ell\sin\theta}\dd\theta},
\end{align}
so that
\begin{align}
\xi &= \partial_t,\quad \eta = \dd t + h \dd z + \bar{h} \dd\bar{z},\quad h = \frac{\ell}{2z}(1-\cos\theta),\quad \bar{h}=\frac{\ell}{2\bar{z}}(1-\cos\theta).
\end{align}
In these coordinates, we have (assuming the ordering $(t,z,\bar{z})$):
\begin{equation}
\Phi^\mu{}_\nu = \left(\begin{matrix}
0 & -ih & i\bar{h}\\
0 & i & 0\\
0 & 0 & -i
\end{matrix} \right).
\end{equation}
The THF data flows as we start changing $f(\theta)$ and $\varepsilon$. In fact, it was already argued in \cite{Closset:2013vra} that $\varepsilon$ is the only nontrivial modulus of the THF for the index background, at least in the round sphere case, $f(\theta)={\rm const}$. Therefore, we only need to show that the deformations of $f(\theta)$ are trivial. Once this is established, we conclude that the partition function is independent of the squashing profile $f(\theta)$, and only depends on the THF modulus $\varepsilon$ (as well as the flavor background parameters).

Let us perform a variation
\begin{equation}
f(\theta) \mapsto f(\theta) + \delta f(\theta).
\end{equation}
This results in the following variations of $\xi$ and $\Phi^\mu{}_\nu$:
\begin{align}
\Delta\xi &= -\delta f(\theta) \frac{\sin\theta}{f(\theta)^2}\partial_\theta,\quad
\Delta\Phi = \delta f(\theta)\left\{ \frac{\ell\cos\theta\sin\theta}{f(\theta)^2} \partial_\theta\otimes(\dd\varphi + \varepsilon \dd\alpha) + \frac{\cot\theta}{\ell} \partial_\varphi\otimes\dd\theta \right\}.
\end{align}
We need only certain components in the adapted coordinates, namely:
\begin{align}
\Delta\xi^z= -\frac{\delta f(\theta)}{f(\theta)} \frac{z\cos\theta}{\ell},\quad \Delta\Phi^z{}_{\bar z}=\frac{\delta f(\theta)}{f(\theta)}\cdot \frac{iz}{2\bar{z}}(\cos^2\theta+\cos\theta).
\end{align}
Applying the formalism described in \cite[Section 5.3]{Closset:2013vra},  consider the following form valued in the holomorphic tangent bundle to the three-manifold (in the THF sense):
\begin{equation}
\Theta^z = -2i\Delta\xi^z (\dd t + h\dd z) + (\Delta\Phi^z{}_{\bar z} - i\bar{h}\Delta\xi^z)\dd\bar{z}.
\end{equation}
To see if the deformation of THF by $\delta f(\theta)$ is trivial,\footnote{Triviality of a deformation means that it is equivalent, up to Q-exact terms in the action, to a deformation by some diffeomorphism. To check the triviality explicitly, one would have to find the diffeomorphism, and compose its action with the deformation due to $\delta f(\theta)$. The combined operation then deforms the action by a Q-exact term. We rely on a general theory and do not perform this check explicitly.} one has to check whether the form $\Theta^z$ is $\tilde{\partial}$-exact (it is already $\tilde{\partial}$-closed). Here $\tilde{\partial}$ is the analog of Dolbeault operator on a three-manifold with the THF. This amounts to finding a real vector field $v^\mu$, such that
\begin{equation}
\Delta\xi^z = -\partial_t v^z,\quad \Delta\Phi^z{}_{\bar z}=2i\partial_{\bar z}v^z -i\bar{h}\partial_t v^z,
\end{equation}
and we easily find the following solution for $v^z$:
\begin{equation}
v^z = \frac{z}{\ell} g(\theta),\quad \text{where } g'(\theta) = \frac{\delta f(\theta) \cos\theta}{\sin\theta},\text{ and we choose } g(0)=0.
\end{equation}
Since we are looking for a real vector field on $S^1 \times S^2_b$, we have to choose $v^{\bar z}=\frac{\bar z}{\ell} g(\theta)$. The component $v^t$ is not fixed by the equations, and we simply pick $v^t=-g(\theta)$. Transforming such a vector field into the original coordinates $(\theta, \varphi,\alpha)$, it becomes:
\begin{equation}
\label{diffeo}
v^\mu\partial_\mu = -\frac1{\beta} g(\theta)(\partial_\alpha - \varepsilon\partial_\varphi).
\end{equation}
This is a globally defined vector field, at least if we assume $\delta f(\pi-\theta)=\delta f(\theta)$ (i.e., the squashed sphere is invariant under the reflection off the equator). Indeed, we chose $g(0)=0$, which already guarantees that $v$ is continuous at $\theta=0$, where the $\varphi$ circle shrinks. Another point where the $\varphi$ circle shrinks is $\theta=\pi$, so we should have $g(\pi)=0$. This is also true, as a consequence of $\delta f(\pi-\theta)=\delta f(\theta)$, which implies $g'(\pi-\theta)=-g'(\theta)$, and so indeed $g(\pi)=g(\pi)-g(0)=\int_0^\pi g'(\theta)\dd\theta =0$.

Having found a globally defined vector field that trivializes $\Theta^z$, the theory of deformations of the THF outlined in \cite[Section 5]{Closset:2013vra} implies that our deformation $\delta f(\theta)$ of the sphere shape is cohomologically trivial, and does not affect supersymmetric observables. \footnote{Running the same analysis for a deformation $\delta\varepsilon$ shows that the corresponding vector field does not exist, which confirms that $\varepsilon$ is a modulus of the THF. We do not present it here for brevity, and because we do not directly use this fact.}

One thing we would like to point out is that the vector field \eqref{diffeo} does not have a $\theta$ component, so it is compatible with cutting the sphere in half along the equator. Therefore, it is still true for the half-index geometry that a deformation $\delta f(\theta)$ is, up to Q-exact terms, equivalent to a diffeomorphism. One could only worry about the possible total derivative terms in the action, leading to the boundary terms. We will assume that they do not cause any problems: after a deformation by $\delta f(\theta)$ and a diffeomorphism, we get the original action, the Q-exact deformation, and perhaps a boundary term. The action, in the way we wrote it before, is Q-exact, and thus Q-closed, without any boundary terms. Same is true about the deformation term. Thus the boundary term should be supersymmetric on its own, with the appropriate boundary conditions imposed. We will assume the most standard scenario: the boundary term simply vanishes for the supersymmetric boundary conditions.

\subsubsection{Background vector multiplets: FI terms and masses}\label{sec:backgroundVectors}
The standard index background $S^1\times S^2$ \cite{Imamura:2011su} (or its squashed version elaborated above) does not admit the conventional real mass or real FI parameters, where only scalars in the background vector multiplets are turned on. However, it admits a certain generalization thereof, in which other fields in the background vector multiplets are switched on as well. We provide some details on such solutions now.

Bosonic backgrounds for vector multipelts can be found by solving $\delta\lambda=\delta\bar\lambda=0$. Solutions, subject to a simplifying assumption $\partial_\varphi\sigma^{\rm f}=\partial_\alpha\sigma^{\rm f}=F^{\rm f}_{\varphi\alpha}=0$, are parameterized in terms of a $\theta$-dependent ``mass'' function $m(\theta)$. The background fields must obey the following equations:
\begin{align}
\label{bg_vector}
\sigma^{\rm f} &= m(\theta),\cr
F^{\rm f}_{\theta\varphi}&+i\frac{\ell}{\beta-i\ell\varepsilon} F^{\rm f}_{\theta\alpha}=\frac{\beta}{\beta-i\ell\varepsilon}\frac{\dd}{\dd\theta}\left(-\ell m(\theta)\cos\theta \right),\cr
D^{\rm f}&=\frac{i\beta  m'(\theta)}{(\beta-i\ell\varepsilon)f(\theta)\sin\theta} + \frac{\varepsilon\ell D_\theta(m(\theta)\sin\theta)}{(\beta-i\ell\varepsilon)f(\theta)} - \frac{\cos\theta F^{\rm f}_{\theta\alpha}}{(\beta-i\ell\varepsilon)f(\theta)\sin\theta}.
\end{align}
Of course, the same equations govern the backgrounds for both flavor and topological symmetries, where in the latter case the mass function describes the FI parameters.

An important piece of data is whether the solutions to \eqref{bg_vector} must obey reality conditions or not. We never impose any reality constraints on the auxiliary field $D^{\rm f}$. On the contrary, we choose to keep the mass function $m(\theta)$ real (i.e., real masses and real F.I. parameters are \emph{actually} real). As for the background gauge fields, it may be beneficial to complexify them, and indeed we will have to do so. One rational for such a choice is that, as we will see, all the BPS observables depend meromorphically on the complex linear combination
\begin{equation}
\label{ahol_A}
A^{\rm f}_\alpha - \left(\varepsilon + i\frac{\beta}{\ell}\right)A^{\rm f}_\varphi.
\end{equation}
Thus shifting $A^{\rm f}_\alpha$ and $A^{\rm f}_\varphi$ into the complex domain, while preserving this linear combination, is natural, given that we only study the BPS observables.

\paragraph{Flavor background.} The first equation in \eqref{bg_vector} simply determines $\sigma^{\rm f}=m(\theta)$, and the last equation tells that the complexified $D^{\rm f}$ is written in terms of other fields, without any further implications. Here $m(\theta)$ is valued in $\mathfrak{t}={\rm Lie}(\bfT)$, where $\bfT$ is the maximal torus of 3d $\cN=2$ flavor group. The second equation in \eqref{bg_vector}, though, gives a non-trivial constraint on the background gauge fields. Its content depends on whether we insist on keeping $A_\mu^{\rm f}$ real. If we do, the second equation splits in two real equations:
\begin{align}
F^{\rm f}_{\theta\alpha}&=\varepsilon F^{\rm f}_{\theta\varphi},\cr
F^{\rm f}_{\theta\varphi}&=\frac{\dd}{\dd\theta}\left(-\ell m(\theta)\cos\theta \right),
\end{align}
which are solved by the following Cartan-valued connection:
\begin{align}\label{Sol1}
A^{\rm f}_\theta &=0,\cr
\text{Solution 1:}\quad \quad A^{\rm f}_\varphi &= \ell (m(0) - m(\theta)\cos\theta),\cr
A^{\rm f}_\alpha &= \varepsilon\ell (m(0) - m(\theta)\cos\theta) + a_1,
\end{align}
where $a_1$ is an arbitrary constant flat connection on $S^1$, and the expression for $A^{\rm f}_\varphi$ is written in the patch containing $\theta=0$, ensuring that $A^{\rm f}_\varphi\big|_{\theta=0}=0$, for regularity at $\theta=0$. We will call this the real flavor background. When we have the full index geometry $S^1\times S^2_b$, the real flavor background supports a flux through the two-sphere given by:
\begin{equation}
\int_{S^2_b}F^{\rm f}_{\theta\varphi}\dd\theta\dd\varphi = 2\pi \ell (m(0) + m(\pi)), \quad \text{(where $\ell(m(0)+m(\pi))\in Q^\vee$ is a cocharacter)}
\end{equation}
whereas on the hemisphere, the flux is
\begin{equation}
\int_{HS^2_b}F^{\rm f}_{\theta\varphi}\dd\theta\dd\varphi = 2\pi \ell m(0).
\end{equation}
The full sphere index with flux of global symmetry is known as the generalized superconformal index of Kapustin and Willett \cite{Kapustin:2011jm}. Analogously, we obtain \emph{a generalized half-index} by including a flux $2\pi \ell m(0)$ of global symmetry through the hemisphere. The presence of such flux is also related to the notion of twisted quasimaps, as will be explained later. Thus we have the following chain of relations:
\begin{equation}
\text{Solution 1}\quad \Leftrightarrow\quad \text{Generalized index}\quad \Leftrightarrow\quad \text{Twisted quasimaps}
\end{equation}
The generalized index depends non-trivially on the flux \cite{Kapustin:2011jm}, thus we do not expect complete independence on $m(\theta)$ in this case. We will study dependence on $m(\theta)$ shortly.

If we give up the reality of background flavor connection, we obtain a second independent solution characterized by the equations:
\begin{equation}
F^{\rm f}_{\theta\varphi}=0,\quad F^{\rm f}_{\theta\alpha} = i\beta \frac{\dd}{\dd\theta}\left(m(\theta)\cos\theta \right).
\end{equation}
These are solved by the following Cartan-valued connection:
\begin{equation}\label{Sol2}
\text{Solution 2:}\quad \quad A^{\rm f}_\theta=A^{\rm f}_\varphi=0,\quad A^{\rm f}_\alpha = i\beta m(\theta)\cos\theta + a_2,
\end{equation}
where $a_2$ is again a flat flavor connection on $S^1$, which now can be complex by definition. This second solution will be referred to as the complex flavor background. A general background can be a linear combination of the real and the complex solutions.

Like in the study of Janus interfaces in \cite{Dedushenko:2021mds}, let us see what happens under the perturbation,
\begin{equation}
m(\theta) \mapsto m(\theta) + \delta m(\theta).
\end{equation}
We keep the parameters $a_1$ and $a_2$ constant as we vary $m(\theta)$. The flavor background fields only enter the matter kinetic action, so we are looking at the variation:
\begin{equation}
\delta\cL_{\rm c}=D^\mu\bar\phi i\delta A_\mu^{\rm f}\phi - i\delta A^{\rm f}_\mu\bar\phi D^\mu\phi + 2\bar\phi (\sigma+m)\delta m \phi + i\bar\phi \delta D^{\rm f}\phi + \frac{1-2\Delta}{f(\theta)}\bar\phi i \delta A^{\rm f}_{\hat3} \phi + \bar\psi \gamma^\mu \delta A_\mu^{\rm f}\psi + i\bar\psi\delta m \psi.
\end{equation}
Specializing this to the real background, or Solution 1, the variation can be written as
\begin{equation}
\delta\cL_{\rm c}=\{Q,\Xi\}+ \nabla_\mu S^\mu + \frac{i}{\sin\theta} (\bar\phi \delta m(\theta) F+\bar{F}\delta m(\theta)\phi)+\delta m(0) (\dots),
\end{equation}
where $\Xi=-\ell\left[\bar\phi \bar\epsilon(\gamma^\varphi+\varepsilon\gamma^\alpha)\delta m(\theta)\psi+ \epsilon(\gamma^\varphi+\varepsilon\gamma^\alpha)\bar\psi\delta m(\theta)\phi\right]$, and $S^\mu=(-\frac{\bar\phi\delta m(\theta)\phi}{f(\theta)\sin\theta},0,0)$, while the explicit form of the last term is less important. Note that we work with the supercharge $Q= \hat{Q} + \hat{S}$ here.
The terms containing the auxiliary fields $F$ and $\bar{F}$ are treated in the same way as in the Janus interface analysis of \cite{Dedushenko:2021mds}: It was shown there that one can put the auxiliary fields on shell, after which such terms vanish.\footnote{On-shell $\bar{F}\propto \frac{\partial W}{\partial \phi}$, so one obtains $\frac{\partial W}{\partial \phi} \delta m(\theta)\phi$, which, being a variation of $W$ under the global symmetry, is identically zero.} The remaining terms are not automatically zero, unless we demand that $\delta m(0)=0$, and that $\delta m(\theta)$ also vanishes at the other end. Namely, on the full two-sphere, by the other end we mean the South pole, so in this case we must impose:
\begin{equation}
\text{Full sphere:}\quad \delta m(0)=\delta m(\pi)=0.
\end{equation}
If we are on the hemisphere (with $0\leq\theta\leq\frac{\pi}{2}$), then the total derivative produces a non-trivial boundary term, which must vanish, leading to:
\begin{equation}
\text{Hemisphere:}\quad \delta m(0) = \delta m\left(\frac{\pi}{2}\right)=0.
\end{equation}
Thus we see that we are allowed to wiggle the real mass profile $m(\theta)$, as long as the pole and boundary values are kept constant. Hence indeed, the generalized index only depends on $m(0)$ and $m(\pi)$, while the ``generalized half-index'' depends on $m(0)$ and, possibly, $m(\frac{\pi}{2})$.

Now, specializing the variation $\delta\cL_{\rm c}$ to the complex background \eqref{Sol2}, it becomes:
\begin{equation}
\delta \cL_{\rm c}=\{Q,\Xi\} + \nabla_\mu S^\mu + i\sin\theta (\bar\phi \delta m(\theta)F + \bar{F}\delta m(\theta)\phi),
\end{equation}
where $\Xi=i\left[ \bar\phi \bar\epsilon \gamma_{\hat3}\delta m(\theta)\psi + \epsilon\gamma_{\hat3}\bar\psi \delta m(\theta)\phi \right]$, and $S^\mu=(-\frac{\bar\phi \delta m(\theta)\phi \sin\theta}{f(\theta)},0,0)$. The terms containing the auxiliary fields $F$ and $\bar{F}$ vanish for the same reason as before. The total derivative term $\nabla_\mu S^\mu$ disappears completely on the full sphere, since $\delta m(\theta)\sin^2\theta$ vanishes at the poles. If we are on the hemisphere, however, the total derivative generates a non-trivial boundary term. In order for this boudnary term to vanish and $\delta \cL_c$ remain $Q$-exact, we have to impose:
\begin{equation}
\delta m\left(\frac{\pi}{2} \right)=0.
\end{equation}
So we conclude that while on the full sphere, the Solution 2 is trivial, (i.e., describes a $Q$-exact modification of the massless background,) on the hemisphere, it is not. The supersymmetric observables are independent of the mass profile in the bulk $m(\theta)\big|_{0\leq \theta<\frac{\pi}{2}}$, but may depend on the boundary values $m(\theta=\frac{\pi}{2})$. In general, we do not expect the dependence on $m(\frac{\pi}{2})$ to be too complicated: Near the boundary, our setup reminds the elliptic problem studied in \cite{Dedushenko:2021mds}, and there, the dependence on real masses was piecewise constant.

\paragraph{Topological symmetry background.} The analysis above was done for the flavor symmetry background, but most of it was not specific to flavor symmetries. We may apply the results to topological symmetry background vector multiplets as well. They couple to the rest of fields through the supersymmetric BF term \cite{Brooks:1994nn,Kapustin:1999ha} (see also discussions in \cite{Bullimore:2020jdq,Dedushenko:2021mds}). Via such couplings, the above background solutions lead to $\theta$-dependent FI parameters $\zeta(\theta)$, as well as the 3d lifts of the Theta-angle terms in 2d. Here $\zeta(\theta)$ are valued in ${\rm Lie}(\mathbf{A}')$, where we follow the notation of \cite{Dedushenko:2021mds} for the torus $\mathbf{A}'$ of topological symmetries.

Similar to the flavor background, the Solution 2 for topological vector multiplets affords more flexibility in the variations of the FI parameter $\zeta(\theta)$. The FI term constructed in this way corresponds to the following $Q$-exact expression:
\begin{equation}
\label{QexactFI}
\frac12 Q(\bar\epsilon\gamma_{\hat3}\lambda - \epsilon\gamma_{\hat3}\bar\lambda)=D + \frac{i\cos\theta F_{\theta\varphi}}{\ell f(\theta)\sin\theta} - \frac{i D_\theta(\sigma \sin\theta)}{f(\theta)}. 
\end{equation}
We integrate it against $\zeta(\theta)$ to get the FI term in the Lagrangian, which so far is $Q$-exact. Let us also integrate the last term by parts. If we are on the sphere, no boundary terms are generated, and we simply obtain:
\begin{equation}\label{FISol2}
\cL_{\rm FI}= \Tr\left[i\zeta(\theta) D - \zeta(\theta)\frac{\cos\theta F_{\theta\varphi}}{\ell f(\theta)\sin\theta} - (\zeta'(\theta)\sin\theta + \zeta(\theta)\cos\theta)\frac{\sigma}{f(\theta)}\right].
\end{equation}
This is the FI term induced from the Solution 2 (with $a_2=0$) that we need. Again, it is $Q$-exact and so trivial on the full sphere. On the hemisphere (or on an infinite cigar, see below), this term may become nontrivial due to the boundary effect, essentially captured by the boundary term $\propto \zeta \sigma$ generated in the intergation by parts. Of course, the deformation 
\begin{equation}
\zeta(\theta) \mapsto \zeta(\theta) + \delta\zeta(\theta)
\end{equation}
is still $Q$-exact, as long as it vanishes along the boundary, $\delta\zeta\left(\frac{\pi}{2}\right)=0$. This follows trivially from the above discussion: Integrating by parts, we make the deformation manifestly $Q$-exact, due to \eqref{QexactFI}.

Finally, recall that the Solution 2 includes an arbitrary complex flat connection $a_2$ on $S^1$. Such a flat connection for the topological symmetry, via the BF coupling, leads to the Theta-like term in the 3d action:
\begin{equation}\label{Theta3D}
\frac{i\Theta}{4\pi^2} \Tr\int F_{\theta\varphi}\dd\theta\dd\varphi\dd\alpha,
\end{equation}
where $\Theta=2\pi a_2$.

If we chose the Solution 1 for the background topological vector multiplets, the corresponding FI terms would originate from the following $Q$-exact expression:
\begin{equation}
\frac{i\ell}{2}Q\left[\bar\epsilon(\gamma^\varphi+\varepsilon\gamma^\alpha)\lambda -\epsilon(\gamma^\varphi+\varepsilon\gamma^\alpha)\bar\lambda  \right]=D + \frac{\cos\theta (F_{\theta\alpha}-\varepsilon F_{\theta\varphi})}{\beta f(\theta)\sin\theta} - \frac{iD_\theta\sigma}{f(\theta)\sin\theta}.
\end{equation}
The corresponding FI parameter $\zeta(\theta)$ is slightly more rigid, -- it cannot be varied at the poles, -- and like in the flavor case, it corresponds to the possibility of turning on topological fluxes in the generalized superconformal index of \cite{Kapustin:2011jm}.

\subsection{Gomis-Lee stretch}\label{sec:stretch}
Since we have established the independence on $f(\theta)$, we consider stretching the sphere that is part of the $S^1\times S^2_b$ geometry into a very long sausage, like in \cite{Gomis:2012wy}. This may be done with any convenient choice of $f(\theta)$, and we might as well use \eqref{f_of_theta}, in which case the stretch, in terms of the squashing parameter $b=\ell/\tilde{\ell}$, is equivalent to the limit
\begin{equation}\label{GL_stretch}
b\to 0,\quad \text{which we take as } \tilde{\ell}\to \infty,\ \ell={\rm const}.
\end{equation}
As a result, we obtain a very long flat region near the equator $\theta=\frac{\pi}{2}$ of radius $\ell$, which looks like $\bE_\tau\times\R$, where $\bE_\tau$ is the elliptic curve of complex structure $\tau=\varepsilon+ i\frac{\beta}{\ell}$, and where the SUSY parameters, corresponding to $\epsilon_o={1\choose 0}$ and $\bar\epsilon_o={0\choose 1}$, become:
\begin{align}
\label{nearEq}
\epsilon\big|_{\theta=\pi/2} &= \frac1{\sqrt 2}e^{\frac{i}{2}\varphi}{1\choose 1},\cr
\bar\epsilon\big|_{\theta=\pi/2} &= \frac1{\sqrt 2} e^{-\frac{i}{2}\varphi}{1\choose 1}.
\end{align}
Since in the $b\to0$ limit, $\frac{\ell}{f(\pi/2)}\to0$, we see from \eqref{Rbg} that there is an R-symmetry holonomy along the $\varphi$-circle (which is the equator of the sphere) given by:
\begin{equation}
A^{(R)}\approx \frac12 \dd\varphi.
\end{equation}
An R-symmetry gauge transformation can be used to get rid of this holonomy, at the same time removing the factors of $e^{\pm\frac{i}{2}\varphi}$ from \eqref{nearEq}, so we end up with $\epsilon=\bar\epsilon=\frac1{\sqrt 2}{1\choose 1}$ in the near-equator region. The corresponding supercharges $\hat{Q}$ and $\hat{S}$, after careful comparison with the results in \cite[Section 3.1]{Dedushenko:2021mds}, are matched with the supercharges $\cQ_B^\dagger$ and $\cQ_B$, and $Q=\hat{Q} + \hat{S}$ matches the main ``elliptic'' supercharge $\cQ$ of that reference. Its key property (which is why we call it elliptic) is that $\cQ^2 \propto \partial_\alpha - \left(\varepsilon + i\frac{\beta}{\ell}\right)\partial_\varphi$ gives the anti-holomorphic derivative along the elliptic curve, identified with the equator in our index geometry. In the near-equator region, we therefore find the elliptic problem corresponding to $\cQ$. Notice that, because of the R-symmetry gauge transformation that we have done, this relation is only precise when the matter R-charges are integers. If some chiral multiplets have non-integer R-charges, they will end up in the \emph{twisted} sector after the R-gauge transformation, so we will have to be careful about this. We will discuss the choice of R-charge and related matters, such as twisted quasimaps, in Section \ref{sec:quasimaps}.

The global symmetry backgrounds discussed earlier become especially simple near $\theta=\frac{\pi}{2}$. Both Solution 1 and Solution 2 lead to a certain mass profile $m(\theta)$ near the equator, and a completely independent flat connection on the elliptic curve. This is precisely the setting of reference \cite{Dedushenko:2021mds}, where flat connections for flavor symmetries play role of the elliptic equivariant parameters, flat connections for topological symmetries are the so-called ``K{\"a}hler'' parameters, and both masses and FI parameters can vary in the direction transverse to the elliptic curve.

Now let us look at the region close to the pole at $\theta=0$. As we perform the stretch \eqref{GL_stretch}, a small region $0\leq \theta <\theta_0$ is deformed into a thin but long cigar. It is effectively $A$-twisted, and $\hat{Q}+\hat{S}$ matches the 3d A-twist supercharge $\cQ_A$ there. In this region, we make connection with the K-theoretic problem. The region near the other pole at $\theta=\pi$ is likewise stretched into an anti-$A$-twisted, or simply $\bar{A}$-twisted cigar, with $\hat{Q}+\hat{S}$ becoming the conjugate supercharge there. This is a direct 3d analog of the 2d story discussed in \cite{Gomis:2012wy}, motivated by the $tt^*$ geometry \cite{Cecotti:1991me,Cecotti:2013mba}. More precisely, we first write the SUSY parameters near the pole $\theta=0$:
\begin{align}
\epsilon\big|_{\theta=0} &=  e^{\frac{i}{2}\varphi}{1\choose 0},\cr
\bar\epsilon\big|_{\theta=0} &= e^{-\frac{i}{2}\varphi}{0\choose 1},
\end{align}
and find that there is a coordinate singularity there, due to the factors of $e^{\pm\frac{i}{2}\varphi}$. Indeed, the patch and the local frame $e^{\hat a}$ that we have been using so far are only well defined away from the poles. Near the pole, we have to perform a local Lorentz rotation by the angle $\varphi$ to define an alternative local frame
\begin{align}
e^{\tilde 1} &= f(\theta)\cos\varphi \dd\theta - \ell \sin\theta \sin\varphi(\dd\varphi+\varepsilon\dd\alpha),\cr
e^{\tilde 2} &= f(\theta)\sin\varphi \dd\theta + \ell\sin\theta\cos\varphi(\dd\varphi+\varepsilon\dd\alpha),\cr
e^{\tilde 3} &=\beta\dd\alpha,
\end{align}
which is well-defined at $\theta=0$. In this basis, $\epsilon\big|_{\theta=0}={1\choose 0}$ and $\bar\epsilon\big|_{\theta=0}={0\choose 1}$, which correspond to $\hat{Q}=q_-$ and $\hat{S}=\bar{q}_+$ (in the language of \cite[Section 3.1]{Dedushenko:2021mds}), and the total preserved supercharge $\hat{Q}+\hat{S}=\cQ_A$ is indeed the 3d A-twist supercharge. The Spin-connection in this local frame has the following form:
\begin{equation}
\omega^{\hat1\hat2}=\left(1 - \frac{\ell\cos\theta}{f(\theta)} \right)\dd\varphi - \frac{\varepsilon\ell \cos\theta}{f(\theta)}\dd\alpha.
\end{equation}
Let us compare it to the R-symmetry gauge field $A^{(R)}=\frac12 \left(1-\frac{\ell}{f(\theta)} \right)\dd\varphi + \frac{i\beta-\varepsilon\ell}{2f(\theta)}\dd\alpha$. If $\theta$ is very small, $\cos\theta\approx 1$, so $A^{(R)}_\varphi \approx \frac12 \omega^{\hat1 \hat2}_\varphi$. This is the relation necessary for the A-twist along the two dimensional subspace, which in our case is a small neighborhood of the North pole of $S^2_b$. It is approximately obeyed in some small region of angles $\theta< \theta_0$, and becomes exact as $\theta_0\to0$. At fixed $b$, this would imply that the region with the precise A-twist is vanishingly small. Recall, however, that we also take the $b\to 0$ limit. Therefore, we propose that we take these two limits simultaneously, such that $\frac{\theta_0}{b}\to\infty$, that is $b$ tends to $0$ much faster than $\theta_0$. In this case, 
\begin{equation}
f(\theta_0)^2 = \tilde{\ell}^2\sin^2\theta_0 + \ell^2\cos^2\theta_0 \approx \ell^2 (\frac{\theta_0^2}{b^2} + 1) \to \infty,
\end{equation}
showing that the length of the A-twisted tip, $\int_0^{\theta_0}f(\theta)\dd\theta$, can become arbitrarily large. The radius of this tip, however, is really small: It is long but thin. We can illustrate the results of this discussion with the following figure:
\begin{figure}[h]
	\centering
	\includegraphics[scale=0.9]{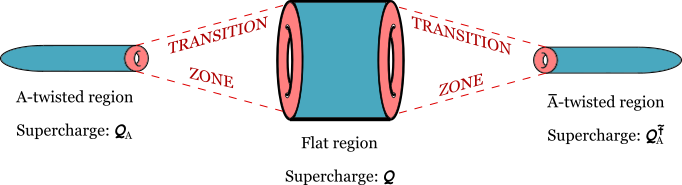}
	\caption{Elongated $S^1\times S^2_b$ geometry. The regions near the North and South poles become the A-twisted and anti-A-twisted cigar geometries, respectively. The near-equator region has the flat $\bE_\tau\times\R$ geometry, with $\tau=\varepsilon+i\frac{\beta}{\ell}$, and the preserved supercharge is the elliptic supercharge $\cQ$ that underlies the relation to elliptic cohomology of \cite{Dedushenko:2021mds}.}
\end{figure}\\

Let us now cut the $S^1\times S^2_b$ geometry along the equator at $\theta=\pi/2$, to obtain a \emph{squashed half-index} geometry $S^1\times HS^2_b$. At the boundary, we impose some supersymmetric boundary conditions that are equivalent to ``capping off'' the half-index geometry with a vacuum state from $\cH[\bE_\tau]$. We do it in an $\cN=(0,2)$ invariant way in the GLSM (i.e., it is a 3d B-brane), because the supercharges $\hat{Q}$ and $\hat{S}$ form the 2d $\cN=(0,2)$ algebra along the boundary. In principle, however, the minimal necessary amount of SUSY preserved by the boundary is $\cN=(0,1)$, because $Q$ (which is used for localization) becomes precisely the $\cN=(0,1)$ supercharge at the boundary. In any case, this is the supersymmetric boundary condition, which produces a boundary state that is $Q$-cohomologous to the chosen vacuum. We illustrate this setup in the following picture:
\begin{figure}[h]
	\centering
	\includegraphics[scale=0.9]{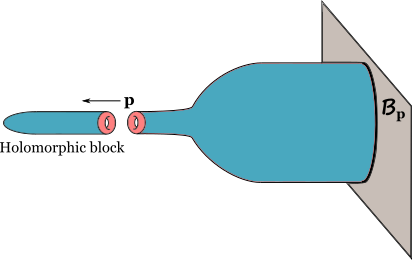}
	\caption{A boundary condition $\cB_p$ creates a boundary state cohomologous to the vacuum $p$. The geometry close to the tip looks like that of a holomorphic block \cite{Beem:2012mb}, with the vacuum $p$ at infinity. The transition zone between the A-twisted cigar and the flat cylinder has a bottle shape in this picture.}\label{fig:half_index_squashed}
\end{figure}\\

This does look very similar to the holomorphic block of \cite{Beem:2012mb}, as the setup computes an overlap $\langle C|p\rangle$, where $\langle C|$ represents the state created by the A-twisted cigar geometry $C\times_\varepsilon S^1$, and $|p\rangle$ denotes the supersymmetric ground state from $\cH[\bE_\tau]$. However, the normalization of $|p\rangle$ does not have to match \cite{Beem:2012mb}, because the long interpolating region that connects the cigar to the boundary $\cB_p$ can contribute an overall factor (and besides that, the boundary itself introduces some normalization factor). Thus $\langle C|p\rangle$ might differ from the holomorphic block by a multiplicative factor. We do know, however, that it matches the half-index with the $\cB_p$ boundary conditions precisely.

Finally, let us mention what happens to our global symmetry backgrounds in the A-twisted cigar region. By definition, the approximation $\cos\theta\approx 1$ is valid there. It is enough to assume that $m(\theta)$ remains constant in that region. Then the background connections both for the Solution 1 \eqref{Sol1} and the Solution 2 \eqref{Sol2} simply become flat connections along the $S^1$ direction. The flavor background in the A-twisted region thus has a very simple form:
\begin{equation}
\sigma^{\rm f}=m={\rm const},\quad A^{\rm f}_\alpha = {\rm const}.
\end{equation}
This background is of course known in the literature on the 3d A-twist. We would like to stress again that without the topological twist, the real mass background cannot be this simple and must include the background gauge field strength in \eqref{Sol1} and \eqref{Sol2}.

The same is observed for the FI parameters in the A-twisted region, which appear in the following simplified Lagrangian term:
\begin{equation}
S_{\rm FI}=i\Tr \int \dd^3 x\,i\zeta D - \Tr\int \beta\zeta F_{\theta\varphi}\, \dd\theta\dd\varphi\dd\alpha,
\end{equation}
where each summand is separately supersymmetric. The first term $\zeta D$ is the standard FI deformation, while the second looks just like the 3d Theta-angle \eqref{Theta3D}. We can combine them, so that for each $U(1)$ factor of the gauge group, we get
\begin{equation}
\left(\frac{i\Theta}{4\pi^2} - \beta\zeta\right) \int F_{\theta\varphi}\, \dd\theta\dd\varphi\dd\alpha.
\end{equation}
Said differently, $\zeta$ shifts the flat background connection $\Theta$ for topological symmetry. Again, such FI terms agree with the literature on the 3d A-twist.

We can easily obtain the full description of the A-twisted region $\theta<\theta_0$. One simply drops various terms in the action and in the SUSY variations that vanish in the limit \eqref{GL_stretch}, such as those proportional to $\frac1{f(\theta)}$. The result is presented in the Appendix \ref{app:Atwist}, and is of course in complete agreement with the A-twisted 3d $\cN=2$ theories \cite{Beem:2012mb,Nekrasov:2014xaa,Benini:2015noa,Benini:2016hjo,Closset:2016arn,Closset:2017zgf}.

\section{Vertex/vortex function}\label{sec:vertexF}
In this section, we focus on the A-twisted index on $S^1\times_\varepsilon C$ ($C$ is the cigar that is eventually compactified to $\C\bP^1$), and clarify some details on its enumerative meaning. Let us drive parameters of our theory into the Higgs phase (assuming it exists), by switching off all masses and turning on generic large (real) FI parameters. We also switch on the background flat connections, both for the flavor torus $\bfT$ and the topological torus $\mathbf{A}'$:
\begin{align}
\label{H_phase}
m(\theta) &= 0,\quad A^{\rm f}=\frac{\tilde{a}}{2\pi} \dd\alpha,\cr
\zeta(\theta) &= \zeta,\quad A^{\rm top} = \frac{\tilde\xi}{2\pi}\dd\alpha.
\end{align}
This background computes certain index that counts BPS states in the Hilbert space $\cH[C]$ associated to our 3d theory placed on
\begin{equation}
\R\times C,
\end{equation}
with the topological twist along $C$, and $\R\ni t$ being the time direction. The localization equations that determine the BPS states in such a 3d A-twisted gauge theory (using the Higgs branch localization, see \cite{Bullimore:2018yyb,Bullimore:2018jlp} and Appendix \ref{app:LocDef}) are:
\begin{align}
\label{Abps1}
D_{\bar w}\phi&=0,\quad \star_C F = -e^2\mu_\R,\cr
\sigma\phi&=0,\quad D_\mu\sigma=0,\quad \partial W=0,\cr
D_t\phi&=0,\quad F_{\varphi t}=0,\quad F_{\theta t}=0,
\end{align}
where $w$ is a complex coordinate on $C$, the Hodge star $\star_C$ acts along $C$, and the real moment map $\mu_\R$ includes the FI shift, roughly $\mu_\R^a=\phi t^a\bar{\phi}-\zeta^a$. We fix a choice of massive vacuum $p$ at the infinity of $C$, assuming that both $\partial W$ and $W$ vanish there.\footnote{This is needed to infer $D_{\bar w}\phi=\partial W=0$ from the term $|D_{\bar w}\phi-e^{i\nu}\bar{\partial W}|^2$ in the localizing action.} Then all fields asymptote to their vevs in the vacuum $p$, and we can compactify the cigar to a projective plane,
\begin{equation}
\bar{C}=\C\mathbb{P}^1\quad \text{(at $[\infty]\in\C\mathbb{P}^1$, fix the vacuum $p$.)}
\end{equation}

Equations in the first line of \eqref{Abps1} are the familiar vortex equations, which also play important role in two dimensions \cite{Morrison:1994fr, Losev:1999nt, Hori:2000kt,Shadchin:2006yz}. Equations in the second line impose restrictions on the geometry of target in which vortices live. Equations in the third line imply that solutions are invariant under the time-translations \cite{Baulieu:1997nj}. In other words, they describe static BPS solitons, and we obtain a moduli space of such solitons. We would like to enumerate them, i.e., compute the BPS-counting index, which mathematically is given in terms of the equivariant integration over the moduli space of such BPS solitons. Physically, this is done via replacing $\R\times C$ by $S^1\times_\varepsilon C$, and turning on holonomies as in \eqref{H_phase}. The twist $\varepsilon$ is an equivariant parameter (or fugacity) for the rotations of $C$, and the flavor holonomies $A^{\rm f}$ are equivariant parameters for global symmetries acting on the Higgs branch. The topological holonomies $A^{\rm top}$ have different meaning, at least when the Higgs branch is pure and not acted on by the topological symmetries: They become counting parameters (fugacities) for the Chern classes of gauge bundles, -- called vortex number(s), or degree of the quasimap below. After passing to $S^1\times_\varepsilon C$, the last line in equations \eqref{Abps1} is replaced by:
\begin{equation}
(D_\alpha - \varepsilon D_\varphi)\phi=0,\quad F_{\varphi\alpha}=0,\quad F_{\theta\alpha}-\varepsilon F_{\theta\varphi}=0,
\end{equation}
which, at the level of equivariant integration, results in the equivariant localization. I.e., it restricts integration to the fixed points of global symmetries and the rotational symmetry $\varepsilon$.

\subsection{Quasimap counts and the 3D A-twist}\label{sec:quasimaps}
If the FI parameters $\zeta$ are finite, it is known that the vortex equations naturally appear in the Higgs branch localization \cite{Benini:2012ui,Doroud:2012xw,Fujitsuka:2013fga,Benini:2013yva}.
For compact $C$ the following rough argument imposes bounds on the allowed vortex numbers.
Integrating the equation $ F = e^2\star_C \mu_\R $ from \eqref{Abps1} over $C$, taking into account the vanishing of $\phi$ for fields of negative electric charge, bounds the total flux $\int_{C} F$ by $\propto e^2 {\zeta} {\rm Area}_{C}$, suggesting an effective picture of Abrikosov-Nielsen-Olesen (ANO) type vortices, packed into $C$, with 
each vortex core area of order $\frac{1}{e^2\zeta}$.

The BPS index is insensitive to continuous variations of the magnitude of the FI parameter.  It may depend on the chamber in the K{\"a}hler cone, to which the FI parameter belongs, i.e. there can be wall-crossing.

Let us qualitatively describe the moduli space of solutions to \eqref{Abps1} in the limit $e^2\zeta\to\infty$, in which case the effective vortex core shrinks to zero size, and the bounds on vortex numbers disappear.
The first equation in \eqref{Abps1}, $D_{\bar w}\phi =0$, means that $G_{\mathbb{C}}$-equivalence class of $\phi(z,\bar{z})$ varies over $C$ in a holomorphic manner.
Generically, it means that $C$ is mapped via $\phi$ to the Higgs branch $X_{H}$. However, if the holomorphic section $\phi$ of the matter bundle passes through the unstable locus, e.g. at some point $x \in C$, the map to $X_H$ is not defined. Moreover, the local behavior of a solution to the second equation in \eqref{Abps1} is precisely that of an ANO vortex, of the vanishingly small area. Such point-like vortices were named freckles in \cite{Losev:1999nt}. Outside freckles, the gauge field curvature can be safely neglected in the limit $e^2 \to \infty$.  Then the second equation in \eqref{Abps1} becomes $\mu_\R=0$, implying that $\phi$ describes a holomorphic map from $C$ (with the freckles, or singularities, removed) into the Higgs branch $X_H$, also denoted $X$ below. At the finite set of singular points, the large gauge field is activated, supporting non-zero Chern classes (in addition to the contribution of small gauge fields outside freckles). Geometrically, at such points the equation $\mu_\R=0$ is violated, so the map goes off the Higgs branch $X_H$ into the unstable locus $\mathscr{X}\setminus X_H$ of the moduli stack. Such freckled instanton solutions are now called (stable) holomorphic quasimaps (or twisted quasimaps) into $X_H$ \cite{CIOCANFONTANINE201417,QmapsKim,Okounkov:2015spn}, see also \cite{Gonzalez:2009fu} for the symplectic approach.
The quasimap is, therefore, roughly, a non-linear superposition of a smooth holomorphic map $C \to X_H$, of lower instanton number, and a number of point-like vortices.

In the Higgs regime, most fields in the vector multiplets vanish, except for the gauge fields whose values are determined by the matter fields. So one can characterize solutions simply by $\phi$, a holomorphic section of the matter bundle $\mathbf{M}$ over $C$. It is more convenient to compactify $C$ and, as was previously mentioned, work with $\bar{C}=\C\mathbb{P}^1$. Then a general twisted quasimap is characterized by a section,
\begin{equation}
\phi\in H^0(\C\mathbb{P}^1,\mathbf{M}),
\end{equation}
obeying $\partial W=0$, and $\phi$'s related by gauge transformations are identified. A twisted quasimap is called stable if (and only if) only finitely many points of $C$ (called singularities of the quasimap) land in the unstable locus of $\mathbf{M}$.

If the image of $\phi$ lies entirely within the stable locus, it simply determines a holomorphic map from $\C\bP^1$ into the Higgs branch $X_H$, or more generally, a holomorphic section of a bundle over $\C\bP^1$ with $X_H$ as a fiber. The latter possibility that $X_H$ (or more generally, the moduli stack) is fibered over $\C\bP^1$ is the reason for the notion of \emph{twisted} quasimaps. Physically, twisting here means that $\C\bP^1$ may support background gauge fluxes for global symmetries, which act on the fiber $X\equiv X_H$ and cause the fibration to be nontrivial.

In general, background fluxes are valued in the torus $\bfT$ of flavor symmetries, and we want them to obey the appropriate integrality conditions, such that we indeed obtain a smooth bundle over $\C\bP^1$. Recall that the topological twist also involves turning on flux, -- for the R-symmetry $U(1)_R$ in this case, -- and the R-charges of all fields must be integers to result in a smooth bundle over $\C\bP^1$. If we mix $U(1)_R$ with some $U(1)_F$ flavor symmetry, the topological twist implies that $\C\bP^1$ supports a flux of $U(1)_F$, which further contributes to the ``twisting'' of quasimaps.

An important class of examples are 3d $\cN=4$ theories of cotangent type, especially those leading to the Nakajima quiver varieties. Their matter consists of full hypermultiplets in a complex representation $\cR$ and, from the 3d $\cN=2$ viewpoint, a distinguished adjoint-valued chiral $\Phi$. The condition $\partial W=0$ removes $\Phi$ and leads to the complex moment map constraint $\mu_\C=0$, so one can drop $\Phi\in {\rm adj}$ in the definition of matter bundle:
\begin{equation}
\mathbf{M}=\cM\oplus\bar{\cM}\otimes \hbar^{-1},\quad \text{with } \cM = \cR\otimes\hbar^{-1/2},
\end{equation}
where $\hbar$ is a line bundle for the global symmetry $U(1)_\hbar = {\rm AntiDiag}[U(1)_H\times U(1)_C]$, where $U(1)_H\times U(1)_C\subset SU(2)_H\times SU(2)_C$ is the maximal torus of the $\cN=4$ R-symmetry group. In this case the torus $\bfT$ of $\cN=2$ flavor symmetries factorizes as in \cite{Dedushenko:2021mds,Bullimore:2021rnr}:
\begin{equation}
\bfT = \mathbf{A} \times U(1)_\hbar,
\end{equation}
where $\mathbf{A}$ is the torus of $\cN=4$ flavor symmetries. We will work equivariantly with respect to $\bfT$, denoting the $\mathbf{A}$ equivariant parameters by $x$ and the $U(1)_\hbar$ parameter by $\hbar$, in a slight abuse of notations. The complex bundle $\cR$ for the case of Nakajima varieties has the form
\begin{equation}
\cR = \bigoplus_{i,j}{\rm Hom}(\cV_i,\cV_j)\otimes \cQ_{ij} \oplus \bigoplus_{i}{\rm Hom}(\cW_i, \cV_i),
\end{equation}
where $\cV_i$ are the color bundles (e.g., for a gauge group factor $U(N)$, the corresponding $\cV$ is a rank-$N$ complex vector bundle), $\cW_i$ are the framing (or flavor symmetry) vector bundles, and $\cQ_{ij}$ are the multiplicity bundles.

If we study \emph{untwisted} quasimaps, the bundles $\cW_i$, $\cQ_{ij}$ and $\hbar$ must be trivial, i.e., they are representation spaces recording the actions of the corresponding global symmetries. Also, in the untwisted case the R-symmetry $U(1)_R$ must be chosen in such a way that $\phi$ are scalars on $\C\bP^1$. Since $U(1)_R$ is some combination of $U(1)_H$ and $U(1)_C$, this uniquely identifies
\begin{equation}
U(1)_R = U(1)_C,
\end{equation}
as the hypermultiplet scalars are neutral under the $U(1)_C$, and so remain scalars after the topological twist. This corresponds to the trivial $\hbar$ line bundle. Other choices of $U(1)_R$ are equivalent to turning on a flux for $U(1)_\hbar$, i.e., making the line bundle $\hbar$ nontrivial, leading to twisted quasimaps. For example,
\begin{equation}
U(1)_R=U(1)_H,
\end{equation}
after topological twisting, makes $\phi$ spinors on $\C\bP^1$. A choice $U(1)_R= {\rm Diag}[U(1)_H\times U(1)_C]$ made in \cite{Dedushenko:2021mds} (i.e., $R = \frac12 (R_H+R_C)$) would lead\footnote{We stress that \cite{Dedushenko:2021mds} work with this R-symmetry but do not perform the topological twist.} to $\phi$ valued in the ``square root of spinors'', which is not a smooth bundle on $\C\bP^1$. In such cases, perhaps, one could study \emph{ramified} twisted quasimaps, which are not valued in a smooth bundle over $\C\bP^1$. Rather, one may choose a ramification point, most conveniently $[\infty]\in\C\bP^1$ or $[0]\in\C\bP^1$, around which the bundle has a non-trivial monodromy. Further twistings of quasimaps may be introduced by making bundles $\cW_i$ and $\cQ_{ij}$ nontrivial, i.e., turning on fluxes for other flavor symmetries. The idea of promoting flavor symmetries to topologically non-trivial backgrounds has been explored in twisted supersymmetric gauge theories in \cite{Hyun:1995mb,Losev:1997tp}, more recently  in the context of  the generalized index of Kapustin-Willett \cite{Kapustin:2011jm}. Note that \cite{Okounkov:2015spn,Okounkov:2016sya,Aganagic:2016jmx,Aganagic:2017smx,Okounkov:2020nql} only twist by such $\cN=4$ flavor symmetries, not by $U(1)_\hbar$. We summarize different possibilities as follows:
\begin{table}[h!]
	\centering
	\begin{tabular}{||c|c||} 
		\hline
		Background & Higgs phase solutions \\ [0.5ex] 
		\hline\hline
		$U(1)_R=U(1)_C$; no flavor fluxes & quasimaps \\ 
		\hline
		$U(1)_R=U(1)_C$; integer flavor fluxes & twisted quasimaps in \cite{Okounkov:2015spn} \\ 
		\hline
		General $U(1)_R$ with integer R-charge; integer flavor fluxes & general twisted quasimaps \\
		\hline
		Arbitrary $U(1)_R$; non-integer flavor fluxes & ramified twisted quasimaps  \\ [1ex] 
		\hline
	\end{tabular}
\caption{Index backgrounds counting different types of quasimaps.}
\label{table:quasi}
\end{table}\\
We will mostly work with twisted quasimaps in the sense of \cite{Okounkov:2015spn}, i.e., choose $U(1)_R=U(1)_C$. Only for the questions of mirror symmetry discussed in Section \ref{sec:wallcross} it may be important to also briefly look at $U(1)_R=U(1)_H$.\footnote{Note that the virtual dimension of \eqref{defQM} is $\dim X$ or $0$ in the $U(1)_C$ and $U(1)_H$ cases, respectively \cite{Bullimore:2018jlp}.}

\paragraph{K-theoretic count} Now let us recall how the K-theoretic counting of quasimaps is introduced. One defines
\begin{equation}
\label{defQM}
{\rm QM}(X) = \{ \text{stable twisted quasimaps to } X \}/\sim,
\end{equation}
where the data of gauge bundles (color bundles $\cV_i$ in the quiver case) and section $\phi$ varies in the moduli space, while the curve $\bar{C}$ and background gauge fields (the twisting bundles $\cW_i$, $\cQ_{ij}$ and $\hbar$) are fixed. 
Next we require the image of $[\infty]\in\C\bP^1$ to lie in the stable locus. This is an open condition, i.e., such quasimaps form an open subset:
\begin{equation}
{\rm QM}_{{\rm nonsing },\infty}(X) \subset {\rm QM}(X).
\end{equation}
On such a subset, the evaluation at $[\infty]$ determines a well-defined map
\begin{equation}
{\rm ev}_{\infty}: {\rm QM}_{{\rm nonsing },\infty}(X) \to X.
\end{equation}
K-theoretic invariants of quasimaps are constructed by pushing forward the K-theory classes from ${\rm QM}_{{\rm nonsing },\infty}(X)$ to $X$ via ${\rm ev}_{\infty}$. This faces an obstacle that the latter map is not proper: it maps non-compact subsets to compact ones. Colloquially speaking, the pushforward involves integration over the space of quasimaps ${\rm f} \in {\rm QM}(X)$ obeying ${\rm f}([\infty])=x$ for each fixed $x\in X$. Such a space is non-compact, simply because point-like vortices are allowed to get arbitrarily close to $[\infty]$, but cannot sit exactly at $[\infty]$.

How do we integrate over a non-compact space? A well-known trick is to replace integration by the ``equivariant'' integration with respect to some symmetry,\footnote{Some examples of equivariant integration in the mathematical physics literature: \cite{Witten:1992xu,Givental:1996,Moore:1997dj,Lossev:1997bz,Nekrasov:2002qd}, also \cite{Dedushenko:2017tdw}.  } namely, use the Atiyah-Bott localization formula as a definition of the integral. In our case, it makes sense to consider $\C^\times_q$ (with the equivariant parameter $q$) rotating $\C\bP^1$ that has two fixed points $[0], [\infty]\in\C\bP^1$. Quasimaps that are fixed by $\C^\times_q$ can only have singularities at $[0]$ and $[\infty]$, and in fact only at $[0]$ if we stay within ${\rm QM}_{{\rm nonsing },\infty}(X)$. Thus ${\rm QM}_{{\rm nonsing },\infty}(X)^{\C^\times_q}$ is compact: All singularities of such quasimaps sit at $[0]$, so they cannot cause non-compactness by approaching $[\infty]$. Therefore the $\C^\times_q$-equivariant (or simply $q$-equivariant) pushforward is well-defined. We denote it as $({\rm ev}_\infty)^q_*$.

Then the Vertex function (without descendants) is defined \cite{Okounkov:2015spn} via $q$-equivariant pushforward in the equivariant K-theory:
\begin{equation}
\label{Vertex_def}
V = ({\rm ev}_\infty)^q_* ({\rm QM}_{{\rm nonsing },\infty}(X), \hat{\cO}_{\rm vir}\, \sz^{{\rm deg}}) \in K_{\bfT\times\C^\times_q}(X)_{\rm loc} \otimes \Q[[\sz]],
\end{equation}
where $\hat{\cO}_{\rm vir}$ will be discussed later, and $\sz$ are fugacities counting the degree of quasimaps, -- physically, $\sz$ represents (complexified) flat connections on $S^1$ for the topological symmetry; they count the degree through the 3d BF term \eqref{Theta3D}. Besides $\sz$, the Vertex function also depends on the equivariant parameters, which include $x$ for the $\cN=4$ flavor symmetries, $\hbar$ for the $U(1)_\hbar$, and $q$ for the $\C^\times_q$ automorphism of $\C\bP^1$. Also note that $V$ takes values in the ``localized'' equivariant K-theory $K_{\bfT\times\C^\times_q}(X)_{\rm loc}$, as a result of $q$-equivariant integration, which means that the expression for $V$ may contain equivariant parameters in the denominators. This is quite typical for equivariant integrals. When the integration domain is compact, the answer is polynomial in the equivariant parameters, and if the integration domain is non-compact, the answer has denominators.

As an element of the localized K-theory, $V$ can be decomposed into the fixed point basis, and the coefficient in front of the $p$'th fixed point class is computed by integration over the quasimaps sending $[\infty]$ to the gapped vacuum $p$. Thus it makes sense to consider the space
\begin{equation}
{\rm QM}_{p,\infty}(X)\subset {\rm QM}_{{\rm nonsing },\infty}(X)
\end{equation}
of quasimaps that land in the massive vacuum (i.e., fixed point) $p\in X$ at $[\infty]\in\C\bP^1$. The physical path integral with the vacuum $p$ chosen at infinity localizes precisely to the integral over such quasimaps.

\textbf{Remark.} For some questions, it might be important to distinguish between $C=\C$, $C=\text{cigar}$ and $C=\C\bP^1\setminus [\infty]$, even though from the viewpoint of complex geometry these spaces are indistinguishable. Namely, the mode associated to the quasimap (asymptotic) value at $[\infty]$ is not normalizable (and hence non-dynamical) on $\C$ and on the cigar, but is dynamical on $\C\bP^1\setminus [\infty]$. Additionally, a subleading mode that corresponds to the derivative of quasimap at $[\infty]$ is non-normalizable (leads to the logarithmic IR divergence) on $\C$, but it becomes dynamical and perfectly fine on the cigar and on $\C\bP^1\setminus [\infty]$. Thus the moduli space of quasimaps can be slightly different for $\C$, $\text{cigar}$ and $\C\bP^1\setminus [\infty]$, depending on how we treat these dangerous modes. Such issues may play role in other contexts \cite{Nekrasov:2012xe}. In our case, we either work with $C=\text{cigar}$ or $C=\C\bP^1\setminus [\infty]$, and we fix the quasimap value at $[\infty]$ to be $p\in X$. The remaining modes are normalizable on these spaces, so the issue of dangerous modes does not arise in the current paper. This problem may be important for other, more subtle questions \cite{Hanany:2003hp,Hanany:2004ea} that will be discussed elsewhere.

The factor of $\sz^{{\rm deg}}$ in \eqref{Vertex_def} is interpreted as the contribution of classical action in the SUSY localization:
\begin{equation}
e^{-S_{\rm cl}} = \sz^{{\rm deg}}.
\end{equation}
The sheaf $\hat{\cO}_{\rm vir}$ appearing in the definition \eqref{Vertex_def} captures the one-loop determinant in SUSY localization. It describes the residual fluctuations (i.e., those that do not cancel out) around the locus of quasimaps in the space of fields. The integration over the localization locus is interpreted as the $q$-equivariant pushforward of this sheaf with respect to
\begin{equation}
{\rm ev}_\infty: {\rm QM}_{p,\infty}(X) \to p\in X.
\end{equation}
Let us look closer at the one-loop determinant. For that, we first list fields written in the 3d A-twisted variables (see \cite{Closset:2016arn,Closset:2017zgf,Closset:2019hyt} for the structure of twisted multiplets, and \cite{Bullimore:2018yyb,Bullimore:2018jlp} for similar analyses), specifying bundles that they are valued in:
\begin{align}
\label{Atwist_fields}
\text{Chirals in }\cM&:\quad q, \psi \in\Gamma(\cM),\quad \chi_{\bar z}\in \Gamma(\cM\otimes \bar{K}),\cr
\text{Chirals in }\bar\cM&:\quad \tilde{q}, \tilde{\psi} \in\Gamma(\bar\cM\otimes\hbar^{-1}),\quad \tilde\chi_{\bar z}\in \Gamma(\bar\cM\otimes \bar{K}\otimes\hbar^{-1}),\cr
\text{Adjoint chiral}&: \quad \Phi_z, \psi_z \in \Gamma({\rm adj}\otimes K\otimes \hbar),\quad \chi_{z\bar{z}}\in \Gamma({\rm adj}\otimes K\otimes \bar{K}\otimes \hbar)\cr
\text{Vector}&:\quad A_t,\sigma,\Lambda_0,\tilde\Lambda_0\in\Gamma({\rm adj}),\quad A_z, \Lambda_z \in \Gamma({\rm adj}\otimes K),\quad A_{\bar z}, \tilde\Lambda_{\bar z}\in\Gamma({\rm adj}\otimes \bar{K})\cr
\text{Ghosts}&: \quad c,\tilde{c}\in\Gamma({\rm adj}).
\end{align}
Here each of the bundles is over $\C\bP^1$, and $K$ is the canonical bundle. When varying quasimaps, all these become bundles over $\C\bP^1\times {\rm QM}(X)$. We use the adapted coordinates of the 3d A-twist: $z$ is a holomorphic coordinate on $\C\bP^1$, and $t=\beta\alpha$ -- coordinate in the third, that is $S^1$, direction. A possible gauge-fixing condition is $D^\mu A_\mu=0$, where the covariant derivative $D_\mu$ contains gauge field corresponding to the background quasimap solution, and $A_\mu$ describes fluctuations around such background. Naturally, the ghost action is $\tilde{c} D^\mu D_\mu c$.

Since the 3d A-twist is topological along $\C\bP^1$, the partition function is independent of its size. In particular, the non-zero modes (whose effective 1d mass is determined by the volume of $\C\bP^1$) must all cancel out in the one-loop determinants. It is enough to only enumerate the zero modes. As usual, they are given by holomorphic sections of the respective bundles:
\begin{enumerate}
	\item Gauge field zero modes: $A_t\in H^0({\rm adj})$ and $A_z\in H^0({\rm adj}\otimes K)$. Then $D_z A_{\bar z} + D_{\bar z} A_z=0$ and the gauge-fixing condition, at the level of zero modes, becomes
	\begin{equation}
	D_t A_t=0.
	\end{equation}
	This 1d gauge supplied by conditions at $[\infty]$ eliminates $A_t$ as a dynamical field. The gauge-fixing condition leaves behind a non-universal determinant $(\det_{H^0({\rm adj})} D_t)^{-1}$, which combines with a non-universal Faddeev-Popov determinant $\det_{H^0({\rm adj})} D_t D_t$ generated by the ghosts, to give a universal factor of $\det_{H^0({\rm adj})} D_t$. It cancels against the contribution of $\sigma$ given by $(\det_{H^0({\rm adj})} D_t^2)^{-1/2} = (\det_{H^0({\rm adj})}D_t)^{-1}$. Fermions $\Lambda_0, \tilde\Lambda_0$ give another factor of $\det_{H^0({\rm adj})} D_t$, and fermions $\Lambda_z, \tilde\Lambda_{\bar z}$ produce a factor of $\det_{H^0({\rm adj}\otimes K)} D_t$, which partially cancels $(\det_{H^0({\rm adj}\otimes K)}D_t^2)^{-1}$ generated by the zero modes of $A_z, A_{\bar z}$. Altogether, the gauge multiplet contributes:
	\begin{equation}
	\label{vec1loop}
	\frac{\det_{H^0({\rm adj})}D_t}{\det_{H^0({\rm adj}\otimes K)}D_t}.
	\end{equation}
	\item As is clearly seen from \eqref{Atwist_fields}, each chiral multiplet contains a pair of complex boson and fermion that are sections of the same bundle, call it $E$ for now, and another fermion that is a section of $E\otimes \bar{K}$, where $\bar{K}=K^{-1}$. The former two have zero modes that are holomorphic sections of $E$, they combine and give a factor of $(\det_{H^0(E)} D_t)^{-1}$. The latter fermion has zero modes that are anti-holomorphic sections of $E\otimes\bar{K}$, or equivalently holomorphic sections of $\bar{E}\otimes K$, giving the contribution $\det_{H^0(\bar{E}\otimes K)} D_t$. Using this result for $E={\rm adj}\otimes K\otimes \hbar$, $\cM$ and $\bar\cM\otimes\hbar^{-1}$, we can write the contributions of adjoint chiral $\Phi$ and chirals comprising the hypers as, respectively:
	\begin{equation}
	\label{chir1loop}
	\frac{\det_{H^0({\rm adj}\otimes \hbar^{-1})} D_t}{\det_{H^0({\rm adj}\otimes K\otimes \hbar)} D_t}\times 	\frac{\det_{H^0(\bar\cM\otimes K)} D_t}{\det_{H^0(\cM)} D_t} \times \frac{\det_{H^0(\cR\otimes K\otimes\hbar)} D_t}{\det_{H^0(\bar\cR\otimes\hbar^{-1})} D_t}.
	\end{equation}
\end{enumerate}
The product of \eqref{vec1loop} and \eqref{chir1loop} is the total one-loop determinant. Using Serre duality, we can rewrite it as:
\begin{align}
&\frac{\det_{H^0({\rm adj})}D_t}{\det_{H^1({\rm adj})^*}D_t}\times\frac{\det_{H^0({\rm adj}\otimes \hbar^{-1})} D_t}{\det_{H^1({\rm adj}\otimes \hbar^{-1})^*} D_t}\times 	\frac{\det_{H^1(\cM)^*} D_t}{\det_{H^0(\cM)} D_t} \times \frac{\det_{H^1(\bar\cM \otimes\hbar^{-1})^*} D_t}{\det_{H^0(\bar\cM\otimes \hbar^{-1})} D_t}\cr
= &\frac{\det_{H^0({\rm adj})}D_t}{\det_{H^1({\rm adj})}D_t}\times\frac{\det_{H^0({\rm adj}\otimes \hbar^{-1})} D_t}{\det_{H^1({\rm adj}\otimes \hbar^{-1})} D_t}\times 	\frac{\det_{H^1(\cM)} D_t}{\det_{H^0(\cM)} D_t} \times \frac{\det_{H^1(\bar\cM \otimes\hbar^{-1})} D_t}{\det_{H^0(\bar\cM\otimes \hbar^{-1})} D_t} = \frac1{\det_{T_{\rm vir}}D_t},
\end{align}
where in the second line we used that $\det_E D_t$ and $\det_{E^*} D_t$ only differ by a sign, and that such signs pairwise cancel here. In the final equality, we express the answer as a determinant over fluctuations along the virtual tangent bundle to the space of quasimaps:
\begin{equation}
T_{\rm vir} = H^0(\C\bP^1,\cT) - H^1(\C\bP^1,\cT),\quad \text{where } \cT = \cM + \bar\cM\otimes\hbar^{-1} - {\rm adj} - {\rm adj}\otimes\hbar^{-1},
\end{equation}
where $\cT$ is viewed as a bundle over $\C\bP^1\times {\rm QM}(X)$. This answer is expected, since the $\C\bP^1$ zero modes describe an effective $\cN=(0,2)$ quantum mechanics\footnote{1d $\cN=(0,2)$ multiplets are the circle reduction of 2d $\cN=(0,2)$ multiplets, as opposed to $\cN=(1,1)$.} into the moduli space of quasimaps ${\rm QM}(X)$. Similar quantum mechanics have been explored in the literature both in the 3d $\cN=4$ and more general 3d $\cN=2$ settings \cite{Bullimore:2016hdc,Bullimore:2018yyb,Bullimore:2018jlp,Bullimore:2019qnt,Bullimore:2020nhv,Bullimore:2021auw}, along the lines of pioneering work \cite{Bershadsky:1995vm} on topological reduction.

All the determinants above omit constant modes on $S^1$, which correspond to the overall integration of differential forms over the space of quasimaps. For a chosen direction in $T_{\rm vir}$ with the Chern root $a$, the determinant of $D_t=\partial_t + \frac{a}{2\pi \beta}$ is regularized to
\begin{equation}
\det D_t = \prod_{n\neq 0}\left(\frac{in}{\beta} + \frac{a}{2\pi\beta}\right) \propto \prod_{n=1}^\infty \left(1+\frac{a^2}{4\pi^2 n^2}\right) = \frac{\sinh a/2}{a/2} = \frac{1 - e^{-a}}{a e^{-a/2}}.
\end{equation}
Thus the total inverse determinant can be written as a product over the Chern roots of $T_{\rm vir}$:
\begin{equation}
\frac1{\det_{T_{\rm vir}}D_t} = \prod_a e^{-a/2}\frac{a}{1-e^{-a}} = {\rm Ch}\left(K^{1/2}_{\rm vir}\right){\rm Td}\left({\rm QM}(X)\right),
\end{equation}
where the virtual canonical class is defined as
\begin{equation}
K_{\rm vir} = \det T_{\rm vir}^{-1}.
\end{equation}
Thus the localization answer, with the help of Grothendieck-Riemann-Roch theorem, can be written as a $\C^\times_q$-equivariant pushforward in the $\bfT\times \C^\times_q$-equivariant K-theory:
\begin{equation}
\label{pushfor}
{\rm eq}\int_{{\rm QM}_{p,\infty}(X)} \sz^{\rm deg}\ {\rm Ch}(K^{1/2}_{\rm vir}){\rm Td}({\rm QM}(X)) = {\rm Ch}\left(({\rm ev}_\infty)^q_* (\cO_{\rm vir}\otimes K_{\rm vir}^{1/2} \sz^{\rm deg})\right).
\end{equation}
This almost gives us $\hat\cO_{\rm vir}$, which is defined in \cite{Okounkov:2015spn} as $\cO_{\rm vir}\otimes K_{\rm vir}^{1/2}$ with one additional factor:
\begin{equation}
\label{Ovir}
\hat\cO_{\rm vir} = \cO_{\rm vir}\otimes \left(K_{\rm vir} \otimes \frac{\det \cT^{1/2}\big|_{[\infty]}}{\det \cT^{1/2}\big|_{[0]}}\right)^{\frac12},
\end{equation}
where $\cT^{1/2}$ is a polarization of $\cT$,
\begin{equation}
\cT = \cT^{1/2} + \hbar^{-1} (\cT^{1/2})^\vee,
\end{equation}
chosen there to be
\begin{equation}
\cT^{1/2} = \cM - {\rm adj}.
\end{equation}
Note that again $\cT^{1/2}$ is a class over $\C\bP^1\times {\rm QM}(X)$, so after restrictions to $[0],[\infty]\in\C\bP^1$, we get classes over ${\rm QM}(X)$.

\paragraph{Details on the insertion in \eqref{Ovir}.} What is the meaning of the ratio of determinants in \eqref{Ovir}? It clearly corresponds to an insertion of some observable under the path integral. Observables compatible with the 3d A-twist are complexified Wilson loops wrapping the $S^1$, roughly given by $\Tr {\rm P} \exp i\oint (A_t - i\sigma)\dd t$, which will be discussed in more detail later. Here, insertions of the determinant bundle $\det \cT^{1/2}$ correspond to specific abelian Wilson loops, namely the Wilson loop charged in $\det \cT^{1/2}$ at $[\infty]\in\C\bP^1$ and the Wilson loop charged in $(\det\cT^{1/2})^{-1}$ at $[0]\in\C\bP^1$. Contributions of such Wilson loops almost cancel, leaving only a power of $q$ that measures the Chern class of $\det\cT^{1/2}$. This will be clear from the way Wilson loops contribute in the twisted index discussed later, see also relevant details on the A-twisted background explained in \cite{Beem:2012mb}, namely, why the holonomies of $A_\alpha$ at the two poles of $\C\bP^1$ differ by $q$ to the power of abelian flux. Mathematically, it is explained in \cite{Okounkov:2020nql} that on the $\C^\times_q$-fixed locus, the insertion becomes very simple and, as promised:
\begin{equation}
\label{Det_insert}
\frac{\det \cT^{1/2}\big|_{[\infty]}}{\det \cT^{1/2}\big|_{[0]}}\Bigg|_{{\rm QM}_{\rm nonsing,\infty}(X)^{\C_q^\times}} = q^{-c_1(\cT^{1/2})}.
\end{equation}
Indeed, $\frac{\det \cT^{1/2}\big|_{[\infty]}}{\det \cT^{1/2}\big|_{[0]}}$ is a copy of $\C$ charged only under $\C^\times_q$. The trivializations of $\det\cT^{1/2}$ over the patches of $\C\bP^1$ containing $[0]$ and $[\infty]$ are related by the gluing cocycle $z^{\deg \cT^{1/2}}$. Since $\C^\times_q$ acts by $z\mapsto qz$, this implies \eqref{Det_insert}. In turn \eqref{Det_insert} simply means that such an insertion is equivalent to a shift of the K{\"a}hler variables $\sz$ by half-integral powers of $q$.

To see this, assume that the gauge group center is $U(1)^c$. Then there are $c$ K{\"a}hler variables $\sz_1,\dots,\sz_c$, and the degree of quasimap is $(m_1,\dots,m_c)$, where $m_i\in\Z$ is the first Chern class of the gauge bundle associated to the $i$-th copy of $U(1)$. The degree-counting factor in the Vertex function becomes:
\begin{equation}
\sz^{\deg} = \sz_1^{m_1}\dots \sz_c^{m_c}.
\end{equation}
At the same time $c_1(\cT^{1/2}) = c_1(\cM)$ takes into account multiplicities of the matter representations. As a vector space, $\cM=\cR\otimes\hbar^{-1/2}\cong\C^n$, and there is a $c\times n$ matrix $Q_a{}^i$ of charges of these $n$ complex fields with respect to the center $U(1)^c$. Then if we denote $n_a=\sum_{i=1}^n Q_a{}^i$, the first Chern class (or degree) evaluates to
\begin{equation}
c_1(\cT^{1/2})=m_1 n_1 + \dots + m_c n_c.
\end{equation}
Thus, the insertion \eqref{Det_insert}, appearing under the square root in \eqref{Ovir}, simply shifts each $\sz_a$:
\begin{equation}
\sz_a\mapsto \sz_a q^{-\frac{n_a}{2}},
\end{equation}
compared to the definition derived physically in \eqref{pushfor} that only involves $K_{\rm vir}^{1/2}$ but not the extra insertions at $[0]$ and $[\infty]$.

The motivation of \cite{Okounkov:2015spn} to include such insertions was that $K_{\rm vir} \otimes \frac{\det \cT^{1/2}\big|_{[\infty]}}{\det \cT^{1/2}\big|_{[0]}}$ admits a natural square root that does not include half-integral powers of $q$ (see \cite[Lemma 6.1.4]{Okounkov:2015spn}). This comes at a price of making an extra choice of polarization $\cT^{1/2}$. The physical answer \eqref{pushfor} lacks such an additional insertion, hence it can have half-integral powers of $q$. One may notice that the comparison of Vertex functions defined in the ``math conventions'', i.e., with such an insertion \cite{Aganagic:2016jmx,Aganagic:2017smx}, to physical computations \cite{Bullimore:2020jdq,Okazaki:2020lfy} always involves redefinition of the K{\"a}hler variable by a power of $q$ (and also of $\hbar$ that will be encountered later). Of course the insertion in question is responsible for the discrepancy. The powers of $q$ and $\hbar$ can be traced back to mixed background Chern-Simons terms between the gauge symmetry and the R-symmetry, which are present in the physical analyses.

For later purposes, it is also convenient to define a similar Vertex-like topological index without the insertions of Wilson loops:
\begin{equation}
\label{empty_top_index}
\cI = ({\rm ev}_\infty)^q_* \left({\rm QM}_{{\rm nonsing },\infty}(X), {O}_{\rm vir}\otimes K_{\rm vir}^{1/2}\, \sz^{{\rm deg}} \right).
\end{equation}
Such index may contain half-integer powers of $q$, but it will behave slightly better under the mirror symmetry in Section \ref{sec:wallcross}, by virtue of not containing any loop operators.

\paragraph{Vertex with descendants.} In \cite{Okounkov:2015spn} a generalization called the vertex \emph{with descendants} (or with insertions) is also defined. For that choose a virtual representation of the gauge group
\begin{equation}
\lambda\in K_G({\rm pt}),
\end{equation}
which is simply a tensorial polynomial in $V_i$'s, for example $V_1 - V_3 + V_4\otimes V_5$. One can evaluate $\lambda$ on the fibers of gauge bundles $\cV_i$ over the point $[0]\in\C\bP^1$, that is on $V_i=\cV_i\big|_{[0]}$, and define the vertex with descendants as
\begin{equation}
\langle \lambda| = ({\rm ev}_\infty)^q_* \left({\rm QM}_{{\rm nonsing },\infty}(X), \hat{O}_{\rm vir}\, \sz^{{\rm deg}} \lambda\left(\cV\big|_{[0]}\right) \right).
\end{equation}
In such notations, the vertex itself is $\langle 1|$. We could also be slightly more general and include the flavor symmetry charges,
\begin{equation}
\lambda\in K_{G\times\bfT}({\rm pt}),
\end{equation}
which, however, is not a significant generalization, since the flavor symmetry bundles are kept as constant backgrounds and so do not vary along ${\rm QM}(X)$.

Physically, the insertion $\lambda$ corresponds to a Wilson loop wrapped on $S^1$ and sitting at the tip of $C$. Namely, if the virtual representation is a difference of two actual representations:
\begin{equation}
\lambda =R - R',
\end{equation}
then the insertion is a difference of Wilson loops:
\begin{equation}
\cW_R - \cW_{R'}.
\end{equation}
We further discuss Wilson loops and other BPS insertions in the (half-)index in Section \ref{sec:observables}.

\subsection{Vertex and the half-index}
We argued that the Vertex function $V$ is computed by the topologically twisted index on $S^1\times_\varepsilon \bar{C}$. Meanwhile, the background from Section \ref{sec:squashedIndex} computes the conformal half-index, which naively is a different quantity. In Section \ref{sec:stretch}, however, we saw that the two may be related by the infinite squashing limit, $b=\frac{\ell}{\tilde\ell}\to0$. Namely, a small neighborhood of the pole, $0\leq\theta<\theta_0$, extends into a long and thin approximately A-twisted cigar. If we take the limit $b\to 0$ much faster than $\theta_0\to 0$, then the region $\theta<\theta_0$ becomes an infinitely long and thin A-twisted cigar, which precisely carries a unit of curvature flux and a half-unit of the R-symmetry flux necessary for the topological twist. The rest of geometry slowly interpolates between the A-twisted cigar and the approximately flat near-horizon region.

The contribution of vortices is known to manifest itself in the Higgs branch localization scheme \cite{Benini:2013yva,Fujitsuka:2013fga} (see Appendix \ref{app:LocDef} for localizing deformations and equations on $S^1\times_\varepsilon S^2_b$). In the limit of large FI parameters $\zeta$ and for $\varepsilon=0$, we find point-like vortices (singularities of the quasimaps) sprinkled all over the source $\bar{C}=\C\bP^1$. At $\varepsilon\neq 0$, only the BPS solutions invariant under the $U(1)_q$ rotation of $\C\bP^1$ contribute. Thus the point-like vortices can only sit at the fixed points of $U(1)_q$, and actually only at $[0]$, not at $[\infty]$ since our quasimaps are regular there.

So the BPS locus for half-index (in the Higgs branch localization) includes configurations with point-like vortices sitting at the tip. If we send $\zeta\to\infty$ much faster than $b\to 0$, the vortices remain point-like even as we zoom in at the A-twisted thin cigar $\theta<\theta_0$. (Otherwise, vortices would be smeared along the cigar.) Now let us use the topological invariance of the A-twisted region to deform the cigar into a balloon attached to the hemisphere:
\begin{figure}[h]
	\centering
	\includegraphics[scale=1]{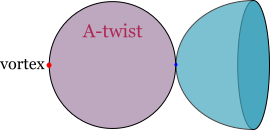}
	\caption{Bubbling off an A-twisted sphere that contains the vortex.}\label{fig:bubble}
\end{figure}\\
To avoid possible confusion among the readers familiar with these matters, note: This is \emph{not} the same as bubbling of $\C\bP^1$ found in the definition of relative quasimaps \cite{Okounkov:2015spn}. We do not alter the definition of quasimaps here. Rather we do a slightly unusual modification of the half-index SUSY background. Since the infinitesimal neighborhood of $\theta=0$ is A-twisted, we ``blow it up'' into an A-twisted bubble that devours the vortex.

A more concrete way to build such a background is by connecting $HS^2$ and $S^2$ by a thin neck of length and radius $\rho$. We assume that $S^1\times_\varepsilon HS^2$ carries the background of Section \ref{sec:squashedIndex}, while $S^1\times_\varepsilon S^2$ carries the A-twisted background. The neck connects and interpolates between the two. We define the SUSY transformations to be $Q$ (of Section \ref{sec:squashedIndex}) on $HS^2$, the A-twist $\cQ_A$ (of Appendix \ref{app:Atwist}) on $S^2$, and interpolate them in some way along the neck. Such combined supersymmetry, --- call it $Q^{(\rho)}$, --- is of course broken, since this is not a supersymmetric background. However, the SUSY-breaking effects are localized inside the neck, and vanish in the $\rho\to0$ limit. Thus in the limit this background becomes supersymmetric. The residual effect of the neck is that $S^1\times_\varepsilon S^2$ and $S^1\times_\varepsilon HS^2$ are glued along $S^1\times \{{\rm pole}\}$, and the fields from the two sides are identified there, see Figure \ref{fig:bubble}. For any field $\phi$ on the right hemisphere and its A-twisted counterpart $\phi^A$ on the left sphere, the identification is simply:
\begin{equation}
\label{touch-glue}
\phi^A\big|_{S^1\times \{{\rm pole}\}} = \phi\big|_{S^1\times\{{\rm pole}\}},
\end{equation}
where ``pole'' refers to the point where the sphere touches the hemisphere. This identification is manifestly supersymmetric, since the $Q$ SUSY variations at the pole $\theta=0$ of $HS^2$ precisely agree with $\cQ_A$.

The condition that vortices may only appear at the opposite pole of the sphere follows from the limiting procedure, but in the final setup of Figure \ref{fig:bubble} should be perhaps viewed as a constraint. What if we now perform the Higgs branch localization of this configuration? On the hemisphere, the BPS equations will only yield the ``Higgs branch solutions'' of reference \cite{Benini:2013yva}. For such solutions, the vector multiplet fields vanish, and the chiral multiplets take constant vevs in the Higgs branch vacuum picked by the boudnary conditions. If we choose some boundary conditions $\cB_p$ picking the fixed point vacuum $p$, then chirals simply localize to this vev on the entire hemisphere. The classical action evaluates to zero on such locus, and there is a non-trivial one-loop determinant
\begin{equation}
Z_{\rm one-loop}[\cB_p],
\end{equation}
which depends on the boundary conditions and will be discussed in a moment. After the fields on $HS^2$ have been localized to the vacuum $p$, they fix via \eqref{touch-glue} the boundary condition at $[\infty]$ for the A-twisted sphere. The A-twisted sphere looks just like the problem considered in Section \ref{sec:quasimaps}, with $\bar{C}=\C\bP^1$ and the vacuum $p$ fixed at $[\infty]\in\C\bP^1$. Such a setup, as we know, computes the $p$'th component of the Vertex $V$ in the fixed-point basis.\footnote{Up to the insertions of abelian Wilson loops that shift $\sz$, which were discussed around eqn. \eqref{Det_insert}.} Thus we conclude that the total partition function is:
\begin{equation}
\label{totalHS2}
\tilde{V}_p = V_p \times Z_{\rm one-loop}[\cB_p] \times e^{\cP[\cB_p]},
\end{equation}
where $e^{\cP[\cB_p]}$ is a contribution from the boundary anomaly that we have not discussed yet.

Perhaps a more traditional explanation of \eqref{totalHS2} would be to say that $Z_{\rm one-loop}[\cB_p] \times e^{\cP[\cB_p]}$ is the perturbative factor in the hemisphere index, while $V_p$ is the non-perturbative K-theoretic vortex partition function counting supersymmetric vortices at the pole.

The notation $\tilde{V}_p$ for \eqref{totalHS2} was not chosen by accident. This is the alternatively normalized Vertex function that appears in \cite{Aganagic:2016jmx} and \cite{Aganagic:2017smx} (though in the latter reference it was simply denoted by $V$, whereas $V$ was denoted ${\rm Vertex}$). More precisely, our $\tilde{V}_p$ for specific choices of boundary conditions matches that from the above references, up to certain overall factors skipped there. To understand this better, let us describe the one-loop determinant explicitly. It is contributed by 3d $\cN=2$ multiplets on the hemisphere, in the background of flat connections. The answer is well-known in the literature \cite{Gadde:2013wq,Yoshida:2014ssa,Dimofte:2017tpi}. Introduce a function (quantum dilogarithm):
\begin{equation}
\upphi(x)\equiv (x;q)_\infty = \prod_{n\geq 0} (1-q^n x).
\end{equation}
A chiral multiplet of R-charge $\Delta$ and flavor weight $x$ (i.e., coupled to the background and/or gauge flat connection $x$) with $\cN=(0,2)$ Dirichlet boundary conditions on the hemisphere contributes
\begin{equation}
\label{Dchir}
e^{-\cE\left(\left(\frac{\Delta}{2}-1\right)\log q + \log x \right)}\upphi(q^{1-\frac{\Delta}{2}} x^{-1})
\end{equation}
in the one-loop determinant, while the one with $\cN=(0,2)$ Neumann gives:
\begin{equation}
\label{Nchir}
e^{\cE\left(-\frac{\Delta}{2}\log q - \log x\right)}\frac1{\upphi(q^{\frac{\Delta}{2}}x)},
\end{equation}
where
\begin{equation}
\cE(x) = -\frac{\log q}{24} - \frac{x}{4} - \frac{x^2}{4\log q}.
\end{equation}
3d $\cN=2$ vector multiplets with $(0,2)$ Dirichlet boundary conditions (in the background of flat gauge connection $s$) contribute:
\begin{equation}
\label{Dvect}
\prod_{\alpha\in{\rm wt(adj)}} e^{\cE\left(-\log\left(q s^\alpha\right)\right)}\frac1{\upphi(q s^\alpha)}.
\end{equation}
The factors involving $\upphi(x)$ in \eqref{Dchir}, \eqref{Nchir} and \eqref{Dvect} precisely capture the contribution to the index that also follows from ``counting letters'' in the free theory \cite{Gadde:2013wq,Dimofte:2017tpi}.
The factors involving $\cE(x)$ follow from the careful zeta-function regularization \cite{Yoshida:2014ssa}, and are usually regarded as part of the boundary anomaly term $e^{\cP[\cB_p]}$ \cite{Bullimore:2020jdq}. Another contribution to $e^{\cP[\cB_p]}$ comes from the Chern-Smions (CS) terms. Even though gauge fields are flat on the hemisphere (recall that we use the Higgs branch localization and hide vortices inside the bubble), they might have non-zero holonomy around $S^1$ in a given vacuum, which contributes to the boudnary term that accompanies the CS term (see \cite{Bullimore:2020jdq}). 

In a 3d $\cN=4$ theory, let us choose $U(1)_R=U(1)_C$ and pick $\cN=(2,2)$ boundary conditions associated to a polarization:
\begin{equation}
T X_H = T^{1/2} + \hbar^{-1} \otimes (T^{1/2})^\vee,
\end{equation}
which we assume to lift to a Lagrangian splitting $L\oplus L^\perp$ of the hypermultiplet matter. This splitting determines $\cN=(2,2)$ boundary conditions on the hypers, such that chirals in $L$ are given the $(0,2)$ Neumann boundary conditions, and $L^\perp$ obey the $(0,2)$ Dirichlet boundary conditions. The adjoint 3d $\cN=2$ chiral multiplet $\Phi$ is given the $(0,2)$ Dirichlet boundary conditions. Finally, we impose Dirichlet boundary conditions on the vector multiplet, with the boundary flat connection $s_p$ corresponding to the vacuum $p$.

Then the one-loop determinant factor is given by
\begin{equation}
Z_{\rm one-loop}[\cB_p] = \prod_{\alpha\in {\rm wt(adj)}}\frac{\upphi(q s_p^\alpha \hbar)}{\upphi(qs_p^\alpha)} \times \prod_{w\in L} \frac{\upphi(q \hbar^{-1/2} s_p^{\rho}x^{w})}{\upphi(\hbar^{1/2} s_p^\rho x^{w})}.
\end{equation}
Following the notations used in \cite{Okounkov:2015spn,Aganagic:2016jmx,Aganagic:2017smx}, we could also write it as:
\begin{equation}
\label{oneLoop_phi}
Z_{\rm one-loop}[\cB_p] = \upphi\left(q\hbar\otimes{\rm adj}_p -q\otimes {\rm adj}_p\right) \upphi\left((q-\hbar)T^{1/2}_p\right),
\end{equation}
where $T^{1/2}_p$ is the restriction of $T^{1/2}$ to the fixed point $p\in X_H$ (i.e., we record weights for the global symmetry torus action on $T^{1/2}_p\subset T_pX_H$). Similarly, ${\rm adj}_p$ records weights $s_p$ of the adjoint gauge bundle corresponding to the constant map to $p\in X_H$.\footnote{In a massive vacuum $p\in X_H$, any transformation from the maxiaml torus of flavor group can be undone by a gauge transformation, which determines the weights $s_p$ as products of the flavor weights $x$ and $\hbar$.}

In \cite{Okounkov:2015spn,Aganagic:2016jmx,Aganagic:2017smx}, the first factor in \eqref{oneLoop_phi} is absent from the definition of $\tilde{V}$ (though \cite{Aganagic:2016jmx} include it in the example of $X_H= T^* \bP^n$, see their equation (94)). For abelian theories, e.g. those with $X_H= T^* \bP^n$, this factor is rather trivial, given by $\left(\frac{\upphi(q\hbar)}{\upphi(q)}\right)^{{\rm rk}(G)}$, which does not affect any further constructions, such as the difference equations obeyed by $\tilde{V}$. In non-abelian theories this factor becomes non-trivial, and our considerations suggest that it must be included. The author of \cite{Okazaki:2020lfy} actually compared the half-index of the $(2,2)$ Dirichlet boundary conditions to the Vertex function $\tilde{V}$ (as defined in \cite{Aganagic:2016jmx,Aganagic:2017smx}) in the simplest nontrivial example of $X_H = T^* \C\bP^1$. He found that (up to the boundary anomaly factor $e^{\cP[\cB_p]}$ to be discussed momentarily) they differ by a similar factor of $\frac{\upphi(q)}{\upphi(q/\hbar)}$, at least if the polarization used in defining the Vertex function $V$ and the polarization used in defining the $(2,2)$ boundary conditions are the same.\footnote{This factor is likewise the contribution of 3d $\cN=4$ vector multiplet, but in the presence of Neumann boundary conditions, which is when the countor integral formulas for $\tilde{V}$ appearing in \cite{Aganagic:2016jmx,Aganagic:2017smx} apply.} We therefore will always include such terms in $\tilde{V}$.

Finally, let us look at the ``boundary anomaly'' term $e^{\cP[\cB_p]}$. It is contributed by the prefactors involving $\cE(x)$ in \eqref{Dchir}, \eqref{Nchir}, and \eqref{Dvect}, and by the additional boundary term associated to the BF coupling between the background topological vector multiplet and the dynamical gauge multiplet. The latter boundary term is given by \cite{Bullimore:2020jdq}:
\begin{equation}
e^{-\frac{\log(\sz)\log(s)}{\log q}}.
\end{equation}
Putting things together, we obtain:
\begin{equation}
\cP[\cB_p]=-\frac{\log(\sz)\log(s_p)}{\log q} - \sum_{(\rho,w)\in L}\frac{\log (\hbar/q)\log(s_p^\rho x^w)}{2\log q} + \frac{\log(\hbar)\log(\hbar q)}{4\log q}\dim(G).
\end{equation}
For quiver theories, $\prod_{(\rho,w)}x^w=1$, while $\prod_{(\rho,w)} s^\rho$ is in the center of gauge group, and in particular couples to the topological symmetry. We thus can write
\begin{equation}
\label{PB_res}
\cP[\cB_p]=-\frac{\log(\sz_\#)\log(s_p)}{\log q} + \frac{\log(\hbar)\log(\hbar q)}{4\log q}\dim(G),
\end{equation}
where following \cite{Aganagic:2016jmx}, we introduced:
\begin{equation}
\sz_\# = \sz (\hbar/q)^{\frac12 {\rm rk}(T^{1/2})}.
\end{equation}
The first term in \eqref{PB_res} gives a contribution to $e^{\cP[\cB_p]}$ that matches $\mathbf{e}(\sz_\#)$ from \cite{Aganagic:2016jmx}, after scaling away $q$ and the appropriate identification of notations. In particular, we use the shorthand where $\log(\sz_\#)$ is an element of $\mathfrak{g}^*$ supported on the center of gauge algebra $\mathfrak{g}$. This means that in the product
\begin{equation}
\log(\sz_\#)\log(s),
\end{equation}
one picks out the center-valued part of $\log(s)$ and contracts it with $\log(\sz_\#)$. Thus we see that $\mathbf{e}(\sz_\#)$ matches the first terms in \eqref{PB_res}. The remaining term is dropped in \cite{Aganagic:2016jmx}, which is safe for their applications to difference equations: The second term in \eqref{PB_res} does not depend on $x$ and $\sz$, and so is inert under their $q$-shifts.

Thus we clearly see that $\tilde{V}$ from \cite{Aganagic:2016jmx,Aganagic:2017smx}, up to certain $(x,\sz)$-independent factors like $e^{\frac{\log(\hbar)\log(\hbar q)}{4\log q}\dim(G)}$ and $\frac{\upphi(q\hbar)}{\upphi(q)}$, and up to abelian Wilson loop insertions \eqref{Det_insert}, matches the half-index of $\cN=(2,2)$ boundary conditions. For this to work, the boundary conditions should be associated to the same polarization that was used in the definition of $V$.

We also introduced the topological index $\cI$ in \eqref{empty_top_index}, which was lacking the abelian Wilson loops \eqref{Det_insert}. It leads to the standard empty half-index, which we denote by:
\begin{equation}
\mathbb{I}_p = \cI_p \times Z_{\rm one-loop}[\cB_p]\times e^{\cP[\cB_p]}.
\end{equation}
This quantity, which is the genuine empty half-index, is closely related to $\tilde{V}_p$. Namely, $\tilde{V}_p$ is equal to $\mathbb{I}_p$ with the K{\"a}hler variable rescaled by a fixed power of $\hbar$ and $q$, as dictated by the insertion \eqref{Det_insert}. $\mathbb{I}_p$ behaves better under the mirror symmetry.

By no means the normalization of $\tilde{V}$ or $\mathbb{I}_p$ is unique or canonical. First, we could use different $(2,2)$ boundary conditions $\cB_p$, associated to different choice of complex polarization of $X_H$. We could also impose Neumann boundary conditions on the vector multiplet (with some boundary matter canceling the anomaly). Second, while the $(2,2)$ boundary conditions have various nice properties \cite{Bullimore:2016nji} and provide a holomorphic Lagrangian basis of boundary states (and in the space of vacua), they are not the only possibility. For example, we could characterize the partition function on $S^1\times_\varepsilon HS^2$ using the fixed point basis, which is realized by the $(0,4)$ Dirichlet boundary conditions. We could also use the space-filling branes with bundles on them (tautological classes) realized via the $(0,4)$ Neumann boundary conditions. This is especially interesting since there is evidence of mirror symmetry between such boundary conditions \cite{Okazaki:2019bok}, however, we will not pursue it further in this work, with an exception of a short comment made towards the end of the next section.

For now, view $\tilde{V}_p$ as a K-theoretic vortex partition function with \emph{some} normalization resulting from a choice of boundary conditions $\cB_p$ (preserving at least $\cN=(0,2)$), or equivalently, the half-index of $\cB_p$. The precise choice of $\cB_p$ is less important and depends on the applications. The main physical content is the state created by quantum fields living on the geometry $S^1\times_\varepsilon HS^2$, while the boundary conditions merely probe this state.

\section{Wall-crossing of the vertex}\label{sec:wallcross}
The Vertex/half-index $\tilde{V}_p$ depends non-trivially on the phase of QFT, and it can experience jumps as the parameters are varied, which, as we will see, plays important role in the mirror symmetry and symplectic duality relations on the Vertex function \cite{Aganagic:2016jmx,Aganagic:2017smx,Rimanyi:2019zyi,Smirnov:2019rmq,Smirnov:2020lhm,Dinkins:2021wvd,Kononov:2020cux}. We focus on real masses $m_\R$ and real FI parameters $\zeta_\R$, and are especially interested in going between the Higgs and Coulomb phases, with the superconformal point sitting in between:
\begin{equation}
\begin{tabular}{|c|}
\hline
Higgs phase\\
$m_\R=0,\ \zeta_\R\gg 0$\\
\hline
\end{tabular} \longleftrightarrow \begin{tabular}{|c|}
\hline
Conformal point\\
$m_\R=0,\ \zeta_\R= 0$\\
\hline
\end{tabular} \longleftrightarrow \begin{tabular}{|c|}
\hline
Coulomb phase\\
$m_\R\gg 0,\ \zeta_\R=0$\\
\hline
\end{tabular}
\end{equation}
The physics and geometry of the space of vacua, as well as the function $\tilde{V}_p$ or $\mathbb{I}_p$, may also depend on the chamber that large $\zeta_\R$ or $m_\R$ take values in.

The $S^1\times_\varepsilon HS^2$ partition function at the conformal point, $m_\R=\zeta_\R=0$, is the original half-index known in the literature \cite{Gadde:2013wq,Dimofte:2017tpi}. Indeed, it does not require turning on somewhat unorthodox position-dependent backgrounds for global symmetries described in Section \ref{sec:backgroundVectors}. In the infinite squashing limit, the state produced at the boundary $\bE_\tau$ of $S^1\times_\varepsilon HS^2_b$ (which has periodic Spin structure, as explained in Section \ref{sec:stretch}) belongs to the Hilbert space $\cH_0[\bE_\tau]$ of the massless theory. This state is determined by the theory and by the choice of R-symmetry that was used in the supersymmetric background on $S^1\times_\varepsilon HS^2_b$. Let us denote such a state by
\begin{equation}
|\mathbb{I}_R\rangle \in \cH_0[\bE_\tau].
\end{equation}
For the two standard choices of 3d $\cN=2$ R-symmetry, $U(1)_C$ and $U(1)_H$, the corresponding states live in different $U(1)_\hbar$ twisted sectors and are denoted:
\begin{equation}
|\mathbb{I}_C\rangle \quad \text{and} \quad  |\mathbb{I}_H\rangle.
\end{equation}
Then the standard half-index of a boundary condition $\cB$ is given by the overlap with the boundary state:
\begin{equation}
\langle \cB|\mathbb{I}_R\rangle.
\end{equation}
Likewise, the state generated by $S^1\times_\varepsilon HS^2_b$ in the Higgs phase will be denoted by
\begin{equation}
|{\rm H}_R\rangle \in \cH_{\rm H}[\bE_\tau],
\end{equation}
and the state in the Coulomb phase is denoted by
\begin{equation}
|{\rm C}_R\rangle \in \cH_{\rm C}[\bE_\tau].
\end{equation}
Again, in both cases we may have $R=C$ or $R=H$.

Interesting enumerative invariants are obtained by taking overlaps of these states with the boundary states admitting geometric interpretations on the Higgs or Coulomb branches. In particular, we now know that the Vertex function of the Higgs branch is the overlap:
\begin{align}
\label{Vert_H}
\mathbb{I}_p &= \langle \cB_p| {\rm H}_C\rangle = \langle p| {\rm H}_C\rangle,\cr
\tilde{V}_p &= \langle \cB_p|\eqref{Det_insert}|{\rm H}_C\rangle = \mathbb{I}_p\big|_{\sz^{\deg} \mapsto q^{-c_1(\cT^{1/2})} \sz^{\deg}},
\end{align}
where we assumed the boundary condition $\cB_p$ to imitate the isolated fixed point vacuum $\langle p|$. One may ask: What do we mean by this, the Higgs branch vacua are clearly not isolated, aren't they? The answer is that the index like $\langle \cB_p|{\rm H}_C\rangle$ involves turning on fugacities (global flat connections on $\bE_\tau$). They play the role of counting parameters in the 3d index, but if we look at the long cigar defining $|{\rm H}_C\rangle$ from far away (in the IR), it becomes a half-line with effective quantum mechanics on it \cite{Beem:2012mb}. The fugacities simply become masses in this quantum mechanics, which gap away the Higgs branch, leaving only the fixed point vacua. Thus it makes sense to talk about $\langle p|$, which mathematically form a fixed point basis in the equivariant elliptic cohomology of the Higgs branch. Such equivariant classes correspond to holomorphic sections of line bundles $\cL_p$ on the elliptic cohomology variety ${\rm E}_T(X_H)$, where the topology of $\cL_p$ is determined by the boundary anomaly of $\cB_p$, with the special choice given by the effective Chern-Simons levels in the massive vacuum $p$ \cite{Dedushenko:2022pem}.

Notice also that in \eqref{Vert_H} we remind the readers about the distinction between $\mathbb{I}_p$ (the empty half-index) and $\tilde{V}_p$ (the half-index with the shift \eqref{Det_insert}).

\subsection{Phase transitions via Janus, and stable envelopes}
In Section \ref{sec:backgroundVectors} we proved that the variations of real mass $m(\theta)$ and real FI $\zeta(\theta)$ profiles are $Q$-exact deformations of the background, so long as we keep their boundary values $m(\pi/2)$ and $\zeta(\pi/2)$ intact.\footnote{Recall that this was the property of Solution 2, whereas Solution 1 also carried nontrivial dependence on $m(0)$ and $\zeta(0)$, corresponding to the generalized index with fluxes.} Thus we can deform $m(\theta)$ and/or $\zeta(\theta)$ to be zero almost everywhere, except for a thin neighborhood of the boundary at $\theta=\pi/2$, where the parameter quickly approaches its boundary value:
\begin{figure}[h]
	\centering
	\includegraphics[scale=0.75]{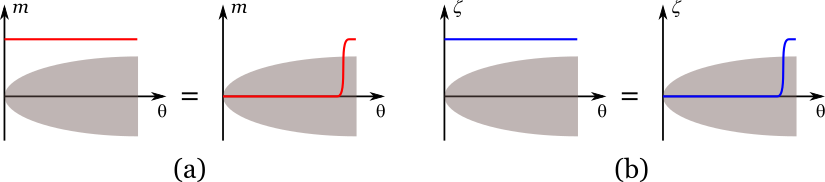}
	\caption{Illustration of how Janus interfaces along the boundary of cigar change the mass (a) and the FI (b) parameters in the bulk.}\label{fig:janus}
\end{figure}\\

The thin region where parameters change is an example of Janus interface of the kind explored in \cite{Dedushenko:2021mds,Bullimore:2021rnr}. Such interfaces can be thought of as operators acting between the spaces $\cH_0[\bE_\tau]$, $\cH_{\rm H}[\bE_\tau]$, and $\cH_{\rm C}[\bE_\tau]$, and they relate the cigar states we introduced earlier:
\begin{equation}
|{\rm H}_R\rangle = \cJ_{\rm FI}(\zeta\leftarrow 0)|\mathbb{I}_R\rangle,\quad |{\rm C}_R\rangle = \cJ_{\rm mass}(m\leftarrow 0)|\mathbb{I}_R\rangle,
\end{equation}
where $\cJ_{\rm FI}(\zeta\leftarrow 0)$ and $\cJ_{\rm mass}(m\leftarrow 0)$ denote the Janus interfaces that change values of the FI parameter from $0$ to $\zeta$, and real mass from $0$ to $m$, respectively. Likewise, we can use them to connect the Higgs and Coulomb phases directly to each other:
\begin{equation}
\label{HC_transition}
|{\rm C}_R\rangle = \cJ_{\rm mass}(m\leftarrow 0)\cJ_{\rm FI}(0\leftarrow\zeta)|{\rm H}_R\rangle,
\end{equation}
where $\cJ_{\rm FI}(0\leftarrow\zeta) = \left(\cJ_{\rm FI}(\zeta\leftarrow 0)\right)^{-1}$.

It was explained in \cite{Dedushenko:2021mds,Bullimore:2021rnr} that $\cJ_{\rm mass}(0\leftarrow m)$ realizes elliptic stable envelopes associated to the Higgs branch, and $\cJ_{\rm FI}(0\leftarrow \zeta)$ does the same for the Coulomb branch. At the level of abstract states, \eqref{HC_transition} in fact provides a physical explanation of the mathematical statement \cite{Aganagic:2016jmx,Rimanyi:2019zyi,Smirnov:2019rmq,Smirnov:2020lhm,Dinkins:2021wvd} that elliptic stable envelopes transform the Vertex function under the 3d mirror symmetry. Let us elaborate this in greater detail.

Under the 3d mirror symmetry, the R-symmetries $SU(2)_H$ and $SU(2)_C$ are swapped. Masses and FI parameters switch their roles, hence the Higgs and Coulomb phases are also swapped. This also includes the identification of Hilbert spaces of some 3d $\cN=4$ theory $T$ and its mirror dual $T^\vee$:
\begin{align}
\cH_0^T[\bE_\tau] \leftrightarrow \cH_0^{T^\vee}[\bE_\tau],\quad \cH_{\rm H}^T[\bE_\tau] \leftrightarrow \cH_{\rm C}^{T^\vee}[\bE_\tau],\quad \cH_{\rm C}^T[\bE_\tau] \leftrightarrow \cH_{\rm H}^{T^\vee}[\bE_\tau].
\end{align}
In particular, the state generated by the $S^1\times_\varepsilon HS^2$ geometry is also identified across the duality, with the flip of phases and R-symmetries taken into account:
\begin{equation}
|\mathbb{I}_C\rangle_T \leftrightarrow |\mathbb{I}_H\rangle_{T^\vee},\quad |{\rm H}_C\rangle_T \leftrightarrow |{\rm C}_H\rangle_{T^\vee},\quad |{\rm C}_C\rangle_T \leftrightarrow |{\rm H}_H\rangle_{T^\vee}.
\end{equation}
Thus if we want to compare the Vertex functions, i.e., the Higgs phase states $|{\rm H}_C\rangle_T$ and $|{\rm H}_C\rangle_{T^\vee}$ of the dual theories, this is the same as comparing the Higgs phase state $|{\rm H}_C\rangle_T$ to the Coulomb phase state $|{\rm C}_H\rangle_T$ of the same theory (up to the swap of parameters and R-symmetries involved in identifying $|{\rm C}_H\rangle_T$ with $|{\rm H}_C\rangle_{T^\vee}$).

To make the comparison more concrete, we must look at the Vertex functions of Higgs branches of $T$ and $T^\vee$, rather than the abstract states. Now we look at the empty half-index $\mathbb{I}_p$ instead of $\tilde{V}_p$:
\begin{equation}
\mathbb{I}_p^T(x,\sz;\hbar) = \langle p|{\rm H}_C\rangle_T,\qquad \mathbb{I}_p^{T^\vee}(x^\vee,\sz^\vee;\hbar^\vee) = \langle p|{\rm H}_C\rangle_{T^\vee}.
\end{equation}
Here we use that the fixed points $p$ on Higgs branches (and thus massive vacua) are identified one-to-one across the mirror symmetry. The half-index $\mathbb{I}_p^T(x,\sz;\hbar)$ depends on the flavor flat connections $x$, topological flat connections $\sz$ and the special flat connection $\hbar$ for $U(1)_\hbar$. The half-index $\mathbb{I}^{T^\vee}_p$ depends on similar parameters of the dual theory: $x^\vee$, $\sz^\vee$ and $\hbar^\vee$. How are these parameters related across the mirror duality? The flavor and topological connections, preserving the full $\cN=4$ SUSY, are simply swapped. Because the $U(1)_\hbar$ generator is proportional to $R_H-R_C$, we expect $\hbar \leftrightarrow \hbar^{-1}$ under the mirror symmetry, due to the swap of $U(1)_H$ and $U(1)_C$. Thus the parameters are identified according to:
\begin{align}
\label{swap_param}
x^\vee &= \sz,\cr
\sz^\vee &= x,\cr
\hbar^\vee &= \frac1{\hbar}.
\end{align}

However, the relationship between functions $\mathbb{I}_p^T(x,\sz;\hbar)$ and $\mathbb{I}_p^{T^\vee}(x^\vee,\sz^\vee;\hbar^\vee)$ is much more interesting than a simple pullback via \eqref{swap_param}. To understand it, perform the mirror symmetry and re-express $\mathbb{I}_p^{T^\vee}$ as:
\begin{equation}
\mathbb{I}_p^{T^\vee}(x^\vee,\sz^\vee;\hbar^\vee) = \langle p|{\rm C}_H\rangle_{T} = \text{function of } (x=\sz^\vee, \sz=x^\vee; \hbar=1/\hbar^\vee),
\end{equation}
where we explicitly indicate that \eqref{swap_param} has been applied. We cannot yet compare $\langle p|{\rm C}_H\rangle_T$ on the right hand side to $\mathbb{I}^T_p=\langle p|{\rm H}_C\rangle_T$ because of two obstructions: (1) they are defined via different R-symmetries; (2) they are computed in different phases.

\paragraph{Changing the R-symmetry.} It is fairly easy to go from $\langle p|{\rm C}_H\rangle_T$ to $\langle p|{\rm C}_C\rangle_T$, i.e. change the R-symmetry used in the SUSY background from $U(1)_H$ to $U(1)_C$. The two differ precisely by $U(1)_\hbar$. In the topologically twisted case, a flux of $U(1)_\hbar$ modifies all spins by the $U(1)_\hbar$ charge, and in the non-twisted round $S^1\times HS^2$, this introduces a shift $q$ of the $U(1)_\hbar$ holonomy along the circle. In either case, this amounts to a replacement
\begin{equation}
\hbar\mapsto q^{-1}\hbar
\end{equation}
in the index which, combined with the mirror identification \eqref{swap_param}, gives $\hbar^\vee \leftrightarrow \frac{q}{\hbar}$. Notice that this is already getting close to the mirror relations on parameters available in the literature, e.g., see \cite{Dinkins:2021wvd}. The additional factors of $\hbar$ and $q$ that appear in the relations between $(x,\sz)$ and $(x^\vee, \sz^\vee)$ in the literature are actually an artifact of using $\tilde{V}_p$, which differs from $\mathbb{I}_p$ by the rescaling of $\sz$. We use $\mathbb{I}_p$ here to avoid this issue.

\paragraph{Changing the phase.}
Now we perform a phase transition, namely, connect $\langle p|{\rm C}_C\rangle_T$ to $\langle p|{\rm H}_C\rangle_T$ using the relation between $|{\rm C}_C\rangle$ and $|{\rm H}_C\rangle$ via Janus:
\begin{align}
\langle p|{\rm C}_C\rangle_T &= \langle p|\cJ_{\rm mass}(m\leftarrow 0)\cJ_{\rm FI}(0\leftarrow \zeta)|{\rm H}_C\rangle_{T}\cr 
&= \sum_r \langle p|\cJ_{\rm mass}(m\leftarrow 0)\cJ_{\rm FI}(0\leftarrow \zeta)|r\rangle\langle r|{\rm H}_C\rangle_{T}\cr
&=\sum_r \langle p|\cJ_{\rm mass}(m\leftarrow 0)\cJ_{\rm FI}(0\leftarrow \zeta)|r\rangle \mathbb{I}_r^T(x,\sz;\hbar),
\end{align}
where we used factorization in the space of SUSY vacua. Since in practice we do computations with boundary conditions rather than the exact SUSY vacua, we also rewrite the latter equality in terms of boundary states, which also fixes a choice of normalization:
\begin{align}
\label{symplectic_dual}
\mathbb{I}_p^{T^\vee}(\sz,x;\frac{q}{\hbar}) =\sum_r \langle \tilde{\cB}_p^\vee|\cJ_{\rm mass}(m\leftarrow 0)\cJ_{\rm FI}(0\leftarrow \zeta)|\cB_r\rangle\frac{1}{\langle \cB_r|\cB_r\rangle} \mathbb{I}_r^T(x,\sz;\hbar),
\end{align}
where $\cB_p$ are the boundary conditions used to define the Vertex/half-index $\mathbb{I}_p^T$ as in \eqref{Vert_H}, and $\tilde\cB_p^\vee$ are the mirror of the boundary conditions $\tilde\cB_p$ used in the theory $T^\vee$ to define $\mathbb{I}_p^{T^\vee}$.

The relations like \eqref{symplectic_dual} between the Vertex functions of the Higgs branches of mirror dual theories are known in the literature \cite{Aganagic:2016jmx,Aganagic:2017smx,Rimanyi:2019zyi,Smirnov:2019rmq,Smirnov:2020lhm,Dinkins:2021wvd}. Clearly, the ``pole-subtraction matrix'' introduced there is:
\begin{equation}
\mathfrak{P}_p{}^r = \langle\tilde{\cB}_p^\vee|\cJ_{\rm mass}(m\leftarrow 0)\cJ_{\rm FI}(0\leftarrow \zeta)|\cB_r\rangle\frac{1}{\langle \cB_r|\cB_r\rangle}.
\end{equation}
We also naturally interpret the equation \eqref{symplectic_dual} as a relation between the Vertex functions of the Higgs and Coulomb branches of the \emph{same} theory:
\begin{align}
\label{symp_dual}
\langle \cB_p|{\rm C}_C\rangle_T =\sum_r \langle {\cB}_p^\vee|\cJ_{\rm mass}(m\leftarrow 0)\cJ_{\rm FI}(0\leftarrow \zeta)|\cB_r\rangle\frac{1}{\langle \cB_r|\cB_r\rangle} \langle \cB_r|{\rm H}_C\rangle_T.
\end{align}
Here both sides are functions of $x,\sz$ and $\hbar$ without any swaps or shifts. This interpretation is physically more natural, as it separates the question of phase transition from the question of identification of parameters in mirror symmetry.
Relation \eqref{symp_dual} can be viewed as part of the Symplectic Duality phenomena, where one connects quantities associated to the Higgs and Coulomb branches of the same theory.

The matrix of coefficients $\langle p|\cJ_{\rm mass}(m\leftarrow 0)\cJ_{\rm FI}(0\leftarrow \zeta)|r\rangle$ describes the vacuum transition amplitude across the Janus interface, and its version $\langle \tilde{\cB}_p^\vee|\cJ_{\rm mass}(m\leftarrow 0)\cJ_{\rm FI}(0\leftarrow \zeta)|\cB_r\rangle$ is computed by the interval partition function. Such objects have been understood to be the elliptic stable envelopes \cite{Dedushenko:2021mds,Bullimore:2021rnr}, up to an extra insertion of $\cJ_{\rm FI}(0\leftarrow \zeta)$ compared to those references. This insertion is, however, very minor, since we saw in Section \ref{sec:backgroundVectors} that the FI term, up to a $Q$-exact Lagrangian, is equivalent to a boundary term $\int_{\bE_\tau}\zeta \sigma \dd^2 x$.

To make the equation \eqref{symplectic_dual} even more concrete, fix a particular basis of boundary conditions suitable for the factorization. If we choose $\{\cB_p\}$ to be the exceptional $(2,2)$ Dirichlet boundary conditions, and $\{\tilde{\cB}_p^\vee\}$ the $(2,2)$ enriched Neumann (mirror to the exceptional Dirichlet of the mirror dual theory), then \cite{Dedushenko:2021mds,Bullimore:2021rnr} show that $\langle\tilde{\cB}_p^\vee|\cJ_{\rm mass}(m\leftarrow 0)|\cB_r\rangle$ computes the elliptic stable envelope. The FI interface $\cJ_{\rm FI}(0\leftarrow\zeta)$ evaluates to the boundary term $\zeta \sigma$, which vanishes at the enriched Neumann boundary, both because $\zeta$ vanishes there and because $\sigma=0$ is enforced by the $(2,2)$ Neumann boundary conditions on the vector multiplet. The remaining bulk FI Lagrangian is $\cQ$-exact. We can deform the FI profile to be $\zeta={\rm const}$ everywhere, because: (1) the boundary term $\zeta\sigma$ at the Dirichlet end is unaffected since $\zeta$ is kept constant there; (2) the boundary term at the Neumann boundary still vanishes due to $\sigma=0$; (3) changing the bulk $\cQ$-exact piece does nothing to the BPS observables. After such deformation, we no longer have the FI Janus, rather $\zeta$ is large and constant everywhere, like in the Higgs phase. We thus find:
\begin{equation}
\mathfrak{P}_p{}^r = \langle\tilde{\cB}_p^\vee|\cJ_{\rm mass}(m\leftarrow 0)|\cB_r\rangle\frac{1}{\langle \cB_r|\cB_r\rangle} = {\rm Stab}^{-1}_{p,r} \times \phi((\hbar-q)T^{1/2}_r)\times e^{\cP},
\end{equation}
where we have ${\rm Stab}^{-1}$ rather than ${\rm Stab}$ because our $\langle\tilde{\cB}_p^\vee|\cJ_{\rm mass}(m\leftarrow 0)|\cB_r\rangle$ acts from the zero mass region to the large mass region, while the stable envelopes conventionally map in the opposite direction. Here the polarization term $\phi((\hbar-q)T^{1/2}_r)$ comes from the normalization factor $\frac{1}{\langle \cB_r|\cB_r\rangle}$, and $e^\cP$ encodes the boundary anomaly contributions. This clearly matches equations from \cite{Aganagic:2016jmx}.

Note that it would also be interesting to perform factorization, and hence obtain similar formulas, in terms of $\cN=(0,4)$ boundary conditions. Exceptional $\cN=(0,4)$ Dirichlet boundary conditions fix point on the Higgs branch (rather than a Lagrangian submanifold), and thus allow to directly construct the fixed point classes. They are expected to be dual to the $(0,4)$ enriched Neumann (see abelian examples in \cite{Okazaki:2019bok}), which in the Higgs phase engineer space-filling branes with bundles. Such branes correspond to the basis of ``tautological classes''.



\section{Supersymmetric observables}\label{sec:observables}
Can we decorate the half-index by insertions consistent with the supersymmetry of the problem? Because $Q^2=(\hat{Q}+\hat{S})^2=\{\hat{Q}, \hat{S}\}$ contains both $D_\alpha$ and $D_\varphi$ for $\theta>0$, the allowed SUSY observables must be surfaces wrapping the $\alpha$ and $\varphi$ circles. The only exception is the point $\theta=0$ at the tip, where the $\varphi$ circle shrinks: There, the BPS insertions must be lines wrapping the $S^1$ parameterized by $\alpha$. Indeed, the SUSY combination at $\theta=0$ is $A_\alpha-i\beta\sigma$, so the BPS observables are complexified Wilson loops,
\begin{equation}
\label{Wilson_0}
W_R = \Tr_R\, {\rm P}e^{i\oint \dd\alpha\, [A_\alpha - i\beta\sigma]\big|_{\theta=0}},
\end{equation}
which are familiar in the 3d A-twisted models \cite{Beem:2012mb,Benini:2015noa,Benini:2016hjo,Closset:2016arn,Closset:2017zgf}. Away from $\theta=0$, we construct surface operators. They couple to the following supersymmetric linear combination, viewed as a complexified anti-holomorphic gauge field:
\begin{equation}
\label{SUSY_A}
\bar\cA_{\bar w}=\frac{i\ell}{2\beta}\left[A_\alpha - \left(\varepsilon + i\frac{\beta}{\ell}\right)A_\varphi -i\beta \sigma\cos\theta\right],\quad Q \bar\cA_{\bar w}=0.
\end{equation}
The corresponding curvature is $Q$-exact:
\begin{equation}
\cF_{\bar{w}\theta}=\partial_{\bar w}A_\theta - \partial_\theta \bar{\cA}_{\bar w} + i[A_\theta, \bar\cA_{\bar w}] = \frac{i\ell}{4\beta\cos\theta} Q(\epsilon\gamma_{\bar{w}\theta}\bar\lambda - \bar\epsilon \gamma_{\bar{w}\theta}\lambda).
\end{equation}
Here we have chosen a complex coordinate
\begin{equation}
w=\varphi + \tau\alpha,\quad \text{where } \tau=\varepsilon + i\frac{\beta}{\ell},
\end{equation}
on the toroidal slice $\mathbb{T}^2_\theta$ at the fixed value of $\theta$. Notice that $A_{\bar w}=\frac1{\bar\tau - \tau}(A_\alpha - \tau A_\varphi)$.

Now we can define a surface operator wrapping $\mathbb{T}^2_\theta$ as a gauge-invariant functional $\Sigma_\theta[\bar\cA]$ of the gauge field $\bar\cA_{\bar w}$ at fixed $\theta$. A well-known way to construct such functionals is via the functional determinants:
\begin{equation}
\Sigma_\theta[\bar\cA] = \frac{\det_{R_1} \bar\nabla}{\det_{R_2} \bar\nabla}.
\end{equation}
Here $R_1$ and $R_2$ are some representations of the 3D gauge group $G$ and, possibly, global symmetry group, which might include 3d global symmetries $G_f\times G_{\rm top}$ and an extra 2d flavor group $F$ of the defect. The covariant derivative $\bar\nabla = \bar\partial + i\bar\cA + i\bar\cG$ contains the complexified gauge field $\bar\cA_{\bar w}$ plus a flat gauge field $\bar\cG_{\bar w}$ for global symmetries. The defect data consist of: (1) defect flavor group $F$; (2) representations $R_1$ and $R_2$ of $G\times G_f\times G_{\rm top}\times F$; (3) flat connection $\cG$ for global symmetries. To ensure that the functional $\Sigma_\theta$ is indeed gauge-invariant, gauge anomalies must cancel. The gauge anomaly of $\det_{R_1}\bar\nabla$ is $\Tr_{R_1} T^a T^b$, where $T^a$ are the generators of $G$, which is seen by realizing $\det_{R_1}\bar\nabla$ as partition function of the $R_1$-valued chiral fermion with the action $S=\int \bar\psi_- \bar\nabla \psi_- \dd^2 w$. Hence the anomaly cancellation condition:
\begin{equation}
\Tr_{R_1} T^a T^b - \Tr_{R_2} T^a T^b =0.
\end{equation}
If we choose to also turn on the background gauge field $\cG$ along the defect, we must ensure cancellation of the mixed gauge-global anomaly as well.

It may be convenient to think of $\Sigma_\theta[\bar\cA]$ as the torus partition function, with periodic boundary conditions, of the $R_1$-valued $bc$ and $R_2$-valued $\beta\gamma$ systems living on the defect:
\begin{equation}
S= \Tr_{R_1} \int b\bar\nabla c + \Tr_{R_2} \int \beta\bar\nabla \gamma.
\end{equation}
The latter is quasi-isomorphic to (the holomorphic twist of) the system of $R_1$-valued Fermi and $R_2$-valued chiral multiplets with $\cN=(0,2)$ SUSY, whose elliptic genus computes $\Sigma_\theta[\bar\cA]$. Examples of $(0,2)$ surface defects, though mostly in 4d theories, can be found in the literature \cite{Maruyoshi:2016caf,Ito:2016fpl}, see also \cite{Gaiotto:2013sma} via breaking of $(2,2)$ defects, and \cite{Gukov:2006jk,Gukov:2008sn,Gaiotto:2009fs,Gaiotto:2012xa,Gadde:2013dda,Nakayama:2011pa,Chen:2023lzq} for variety of other $(2,2)$ and $(4,4)$ examples, including \cite{Chen:2023lzq} where a $(4,4)$ surface defect is broken down to $(0,2)$ by the fugacities. Examples in 3d include \cite{Dedushenko:2021mds,Bullimore:2021rnr}.

More generally, the supercharge $Q$ forms 2d $\cN=(0,1)$ algebra, so $\mathbb{T}^2_\theta$ could support an arbitrary 2d $\cN=(0,1)$ QFT, whose group of flavor symmetries contains a non-anomalous subgroup $G$ gauged by the bulk gauge field $\bar\cA$. The elliptic genera of such systems engineer a large class of examples of $\Sigma_\theta[\bar\cA]$. In any case, $\Sigma_\theta[\bar\cA]$ is simply an elliptic meromorphic function of the gauge-invariant data in $\bar\cA$ (such data, by a complex gauge transformation, is equivalent to a flat connection on $\mathbb{T}^2_\theta$). The precise construction of $\Sigma_\theta[\bar\cA]$ as elliptic genus of some 2d QFT, though instructive, is not very relevant to us.

\subsection{Surface operators and their $\cQ$-exact interpolation}
$\Sigma_\theta[\bar\cA_{\bar w}]$ depends on $\theta$ only implicitly through the $\theta$-dependence of $\cA_{\bar w}$. Then we check that:
\begin{align}
\Sigma_{\theta + \dd\theta}[\bar\cA_{\bar w}] &= \Sigma_\theta[\bar\cA_{\bar w} + \dd\theta \partial_\theta \bar\cA_{\bar w}]=\Sigma_\theta[\bar\cA_{\bar w} + \dd\theta \partial_\theta \bar\cA_{\bar w}- \dd\theta D_{\bar w}A_\theta] = \Sigma_\theta[\bar\cA_{\bar w} + \dd\theta \cF_{\theta\bar{w}}]\cr
&=\Sigma_\theta[\bar\cA_{\bar w}] + \dd\theta \int \frac{\delta \Sigma_\theta}{\delta \bar\cA_{\bar w}} \cF_{\theta\bar{w}}\dd^2 w.
\end{align}
That is:
\begin{equation}
\frac{\dd}{\dd \theta} \Sigma_\theta[\bar\cA_{\bar w}] = \int \frac{\delta \Sigma_\theta}{\delta \bar\cA_{\bar w}} \cF_{\theta\bar{w}}\dd^2 w =\frac{i\ell}{4\beta\cos\theta} Q \int \frac{\delta \Sigma_\theta}{\delta \bar\cA_{\bar w}}(\epsilon\gamma_{\bar{w}\theta}\bar\lambda - \bar\epsilon \gamma_{\bar{w}\theta}\lambda)\dd^2 w.
\end{equation}

Thus the $Q$-cohomology class of $\Sigma_\theta[\bar\cA_{\bar w}]$ does not depend on $\theta$. This conclusion is based on the assumption that the functional $\Sigma_\theta$ does not depend on $\theta$ explicitly. With the definition of $\Sigma_\theta$ via functional determinants as above, this simply means that the torus complex structure $\tau=\varepsilon + i\frac{\beta}{\ell}$ is $\theta$-independent. In other words, we do \emph{not} use the induced metric on $\mathbb{T}^2_\theta$ (which explicitly depends on $\theta$). Rather, we use the constant metric $\dd s^2= |\dd\varphi + \tau\dd\alpha|^2$, which coincides with that of the boundary torus at $\theta=\frac{\pi}{2}$. We could say that the 2d matter engineering $\Sigma_\theta[\bar\cA]$ lives on an auxiliary torus of constant complex structure $\tau$. In particular, this torus does not degenerate even at the point $\theta=0$, where the physical $\varphi$ circle shrinks.

Had we used the induced metric on $\mathbb{T}^2_\theta$ instead, the torus complex structure and thus the partition function $\Sigma_\theta[\bar\cA]$ would explicitly depend on $\theta$. In such a case, $\Sigma_\theta[\bar\cA]$ would not be $Q$-closed at the intermediate values $0<\theta<\frac{\pi}{2}$, interpolating between the BPS ``elliptic'' observable at $\theta=\frac{\pi}{2}$ and the BPS ``trigonometric'' one (the usual complexified Wilson line) at $\theta=0$. This is because the definition of $Q$-closed $\bar\cA_{\bar w}$ in \eqref{SUSY_A} contains a constant complex structure $\tau$ in it. We could also consider $Q$-closed but explicitly $\theta$-dependent $\Sigma_\theta[\bar\cA]$ by allowing the precise choice of underlying elliptic function to vary with $\theta$.

We prefer to work with the $\theta$-independent $\Sigma_\theta[\bar\cA]$, since it is more natural, and it can be moved freely between $\theta=\frac{\pi}{2}$ and $\theta=0$. At the pole, the ``auxiliary torus'' does not degenerate. Instead, due to regularity of the gauge field, $A_\varphi=0$ at $\theta=0$.  At such a point, $\bar\cA_{\bar w}$ turns into a complexified gauge field along the circle:
\begin{equation}
\bar\cA_{\bar w} \big|_{\theta=0} = \frac{i\ell}{2\beta} [A_\alpha - i\beta\sigma]\big|_{\theta=0}.
\end{equation}
The standard $Q$-closed and gauge-invariant observable at $\theta=0$ is the Wilson loop, as was mentioned before:
\begin{equation}
W_R = \Tr_R\, {\rm P}e^{i\oint \dd\alpha\, [A_\alpha - i\beta\sigma]}.
\end{equation}
It is invariant under the complexified gauge transformations, and we can use them to make the complexified gauge field constant and valued in the Cartan:
\begin{equation}
(A_\alpha - i\beta\sigma)^{[g]} = a \in \mathfrak{h}\otimes\C.
\end{equation}
Also when we compute the $S^1\times_\varepsilon HS^2$ partition function via SUSY localization, the gauge multiplet takes values in the localization locus mimicking the Solution 1 of Section \ref{sec:backgroundVectors}. In this case $\bar\cA_{\bar w}$ becomes literally $\theta$-independent, equal to a constant $\frac{i\ell}{2\beta}(A_\alpha - i\beta\sigma)\big|_{\theta=0}=\frac{i\ell}{2\beta} a$ valued in $\mathfrak{h}\otimes\C$, which parameterizes the localization locus.

Then the Wilson loop evaluates to:
\begin{equation}
W_R = \Tr_R\, e^{2\pi i a} = \Tr_R\, s,\quad \text{where } s = e^{2\pi i a} \in \mathbf{H}_\C \subset G_\C.
\end{equation}
Here $\mathbf{H}_\C = (\C^\times)^{{\rm rk}(G)}$ is the complexified maximal torus of the gauge group.

In terms of the same parameter $a\in\mathfrak{h}\otimes\C$, the surface operator now reads:
\begin{equation}
\Sigma_\theta[\bar\cA] = \frac{\det_{R_1} (\bar\partial - \frac{\ell a}{2\beta}-\frac{\ell b}{2\beta})}{\det_{R_2}(\bar\partial - \frac{\ell a}{2\beta}-\frac{\ell b}{2\beta})} = \frac{\det_{R_1} (\partial_\alpha -\tau\partial_\varphi +ia + ib)}{\det_{R_2}(\partial_\alpha -\tau\partial_\varphi +ia + ib)},
\end{equation}
where we also included the flat connection $b$ for global symmetries coupled to the surface operator. This expression is manifestly $\theta$-independent, as $\bar\cA_{\bar w}$ equals $\frac{i\ell a}{2\beta}$ on the localization locus. After taking the determinants, $\Sigma_\theta[\bar\cA]$ becomes:
\begin{equation}
\label{ell_obs}
\frac{\prod_{\lambda\in R_1}\vartheta(e^{2\pi i\langle \lambda, a+b\rangle})}{\prod_{\lambda\in R_2}\vartheta(e^{2\pi i\langle \lambda, a+b\rangle})},
\end{equation}
where
\begin{equation}
\vartheta(x) = (x^{1/2} - x^{-1/2})\prod_{n>0}(1-q^n x)(1-q^n x^{-1}),\quad q=e^{2\pi i \tau}.
\end{equation}
\eqref{ell_obs} is a meromorphic elliptic function of $a=(a_1,\dots, a_r)$, where $a_k\sim a_k+1 \sim a_k+\tau$. These variables parameterize the space $\cE_{\mathbf{H}}/\cW$ of flat gauge connections on the elliptic curve $\C^\times/q^\Z$, where $\cE_{\mathbf{H}} = (\C^\times/q^\Z)\otimes \mathbf{H}_\C$ and $\cW$ is the Weyl group. Equivalently, $\cE_{\mathbf{H}}/\cW$ is the space of complex gauge-equivalence classes of $\bar\cA$, that is, the space of equivalence classes of holomorphic structures on the gauge bundle over $\C^\times/q^\Z$. The latter interpretation is more natural, given that we write our surface operators as $\Sigma_\theta[\bar\cA]$. That $\Sigma_\theta[\bar\cA]$ is indeed an elliptic function of $a$'s follows from the anomaly cancellation, and to be nontrivial it must be meromorphic, not just holomorphic. In its dependence on $b$, it is merely a section, since we do not insist on canceling the 't Hooft anomalies of the defect flavor group $F$.

Note that the Wilson lines $W_R$ alone are ``trigonometric'' functions on $\mathbf{H}_\C/\cW$. In our background, they are supersymmetric only when inserted at the tip of the cigar $\theta=0$. By taking the appropriate ratios of (infinite) products of the Wilson lines, as in \eqref{ell_obs}, we obtain elliptic functions defined earlier. Such elliptic observables can be moved away from the tip to an arbitrary value of $\theta$, without changing the $Q$-cohomology class of the observable.

Let us compare the insertions of Wilson and surface operators in their effect on the localization formulas. It is convenient to compute $\tilde{V}_p$ or $\mathbb{I}_p$ in the basis $\{\sD_p\}$ of exceptional Dirichlet boundary conditions labeled by the isolated vacua $p$. They involve imposing $(2,2)$ Dirichlet boundary conditions on the vector multiplets (or, more generally, $(0,2)$ in $\cN=2$ theories), with the boundary connection $s_p$ corresponding to its classical value in the vacuum $p$. The matter multiplets are given boundary vev according to the vacuum $p$, with the remaining freedom chosen to preserve $(2,2)$ or $(0,2)$. When we compute the half-index in the Coulomb branch localization scheme, it known \cite{Dimofte:2017tpi,Bullimore:2020jdq} that the answer is given by a sum over the flux sectors:
\begin{equation}
\mathbb{I}_p=e^{\cP_p} \sum_B \sz^B Z_{\rm 1-loop}(q^B s_p, x,\hbar),
\end{equation}
where the sum goes over the cocharacters, in particular $B\in\Z$ when $G=U(1)$. The presence of flux $B$ shifts the flat connection $s_p$ by $q^B$ in the one-loop determinant. The latter has a general form explained earlier in this paper.

The insertion of $\Sigma_\theta[\bar\cA]$ modifies the above computation in a fairly obvious way. Because our surface operator is an elliptic function of $s$, the shifts by $q^B$ do nothing to it:
\begin{equation}
\sum_B \sz^B Z_{\rm 1-loop}(q^B s_p, x,\hbar) \Sigma[q^B s_p]= \Sigma[s_p]\sum_B \sz^B Z_{\rm 1-loop}(q^B s_p, x,\hbar).
\end{equation}
Hence the one-point function is simply given by evaluating $\Sigma_\theta[\bar\cA]$ on $s_p$:
\begin{equation}
\langle \Sigma_\theta[\bar\cA] \rangle_p = \frac{\prod_{\lambda\in R_1} \vartheta(e^{2\pi i \langle \lambda, s_p +b\rangle})}{\prod_{\lambda\in R_2}\vartheta(e^{2\pi i \langle \lambda, s_p +b\rangle})},
\end{equation}
where the right-hand side is the elliptic function of $s$ underlying the functional $\Sigma$. We can think of this surface operator as acting on the vertex function, and in the basis of boundary conditions $\{\sD_p\}$, this action is diagonal:
\begin{equation}
\label{Sigma_action}
\Sigma[\bar\cA] \tilde{V}_p = \frac{\prod_{\lambda\in R_1} \vartheta(e^{2\pi i \langle \lambda, s_p +b\rangle})}{\prod_{\lambda\in R_2}\vartheta(e^{2\pi i \langle \lambda, s_p +b\rangle})} \tilde{V}_p.
\end{equation}
Now let us compare this to Wilson loops.

\subsection{Wilson loops and quantum K-theory}
The insertion of a more familiar Wilson loop at the pole $\theta=0$ modifies the $S^1\times_\varepsilon HS^2$ partition function $\mathbb{I}_p$ in a less trivial way, which we already anticipated in Section \ref{sec:quasimaps}. Let us consider the abelian $G=U(1)$ for simplicity. The Wilson loop of charge $k$ is represented by the insertion of $s^k$, which gets shifted by $q^{Bk}$ due to flux. Thus we obtain:
\begin{equation}
\sum_B \sz^B Z_{\rm 1-loop} \times q^{Bk} (s_p)^k = (s_p)^k \sum_B (q^k \sz)^B Z_{\rm 1-loop}.
\end{equation}
In other words, the half-index $\mathbb{I}_p(\sz,x,\hbar)$ with the charge $k$ Wilson loop insertion becomes:
\begin{equation}
(s_p)^k \mathbb{I}_p(q^k \sz, x,\hbar).
\end{equation}
That is, the Wilson loop acting on the half-index is represented by a shift operator:
\begin{equation}
W^k = s^k q^{k \sz\frac{\partial}{\partial \sz}}.
\end{equation}
Generalization to the non-abelian case is not hard. Such shift operator representations of the Wilson loops are well-known in the literature \cite{Jockers:2018sfl,Jockers:2019wjh,Jockers:2019lwe,Jockers:2021omw}. We see that Wilson loops act quite differently from the elliptic surface operators we started with.

One can also use these results to derive relations on the Wilson loops. Below we do this in two examples, deriving the relations of quantum K-theory, with the K{\"a}hler variable $\sz$ playing the role of quantum parameter.

\subsubsection{Examples}

\paragraph{$\C\bP^{L-1}$ model}
Let us start with the 3d $\cN=2$ $U(1)$ gauge theory with $L$ charge-one chiral multiplets and the FI parameter $\zeta>0$, such that at low energies one finds $\C\bP^{L-1}$. We also include a bare CS level $k$. The R-charges of all chirals are zero. We impose the $(0,2)$ Dirichlet boundary conditions both on the vector and chiral multiplets, with the boundary vevs determined by the vacuum $p$ (this model has $L$ vacua, in each only one chiral develops a vev). The gauge field, in particular, receives the boundary value $s=s_p$. The half-index reads:
\begin{equation}
\mathbb{I}_p = \sum_{m\in \Z} e^{\cP_m} q^{\frac12 k m^2} s^{km} \sz^m \prod_{i=1}^L \upphi(q^{1-m}s^{-1} x_i^{-1}),
\end{equation}
where
\begin{equation}
e^{\cP_m}=e^{\frac{\log(\sz) \log(s)}{\log q}-\sum_{i=1}^L \cE[-\log q +\log q^m s x_i]} = q^{\frac{L}{24}}  s^{ -\frac{L}{4}}  \left(\frac{s^2}{q}\right)^{\frac{Lm}{4}}  q^{\frac{L m^2}{4}} e^{\frac{\log(\sz) \log(s)}{\log q}+\sum_i \frac{(\log sx_i)^2}{4\log q}}
\end{equation}
Let us use the difference equation obeyed by $\upphi(x)$,
\begin{equation}
\upphi(x) = (1-x) \upphi(qx).
\end{equation}
Note that the series for $\mathbb{I}_p$ only receives contributions from $m\leq 0$, since for $i=p$, $\upphi(q^{1-m}s^{-1}_p x_p^{-1}) =\upphi(q^{1-m})$, which vanishes for $m\in \Z_{>0}$. By shifting $m\mapsto m+1$, we rewrite $\mathbb{I}_p$ as
\begin{align}
&\sum_m e^{\cP_{m+1}} q^{\frac12 k (m+1)^2} s^{k(m+1)} \sz^{m+1} \prod_{i=1}^L \upphi(q^{-m}s^{-1} x_i^{-1})\cr 
= &\sum_m e^{\cP_m} q^{\frac12 k m^2} s^{km} \sz^m \left\{ \sz s^{k+\frac{L}{2}} q^{km+\frac{Lm}{2}+\frac{k}{2}} \prod_{i=1}^L (1-q^{-m}s^{-1} x_i^{-1})\right\} \prod_{i=1}^L \upphi(q^{1-m}s^{-1} x_i^{-1}).
\end{align}
Recall that $q^m s$ is created by the difference operator $W=s q^{\sz\frac{\partial}{\partial \sz}}$ acting on $\mathbb{I}_p$. Thus the insertion in curly brackets is generated by a difference operator leading to the equation:
\begin{equation}
\left(W^{k+\frac{L}{2}}\prod_{i=1}^L (1-W^{-1} x_i^{-1})-\sz^{-1}q^{-\frac{k}{2}}\right) \tilde{V}_p = 0.
\end{equation}
If we adjust the CS level such that $k+\frac{L}{2}=0$, this is the relation of quantum equivariant K-theory of $\C\bP^{L-1}$, otherwise it is quantum equivariant K-theory with level structure \cite{ruan2019level}. The large radius limit $\zeta\to +\infty$ implies $\sz^{-1} \to 0$, and we recover the relation of classical equivariant K-theory.

\paragraph{$T^* \C\bP^{L-1}$ model}
Now consider 3d $\cN=4$ $U(1)$ gauge theory with $L$ charge-one hypermultiplets and FI parameter $\zeta>0$. This theory has the $T^*\C\bP^{L-1}$ Higgs branch. Imposing $(2,2)$ exceptional Dirichlet boundary conditions for the vacuum $p$, with $s=s_p$, we find the followings half-index:
\begin{equation}
\mathbb{I}_p = \sum_{m\in\Z} e^{\cP_m} \sz^m \prod_{i=1}^L \frac{\upphi(q^{1-m}s^{-1}x_i^{-1})}{\upphi(q^{-m}s^{-1} x_i^{-1}\hbar)},
\end{equation}
where
\begin{equation}
e^{\cP_m}=s^{-\frac{L}{2}}q^{-\frac{Lm}{2}}\hbar^{\frac{L}{4}+\frac{Lm}{2}} e^{\frac{\log(\sz)\log(s)}{\log q} + L\frac{\log(\hbar)\log(s^2\hbar^{-1})}{4\log q}}
\end{equation}
Here the sum is also supported on $m\leq 0$. Shifting $m \to m+1$, we obtain:
\begin{align}
\mathbb{I}_p &= \sum_{m\in\Z} e^{\cP_{m+1}} \sz^{m+1} \prod_{i=1}^L \frac{\upphi(q^{-m}s^{-1}x_i^{-1})}{\upphi(q^{-m-1}s^{-1} x_i^{-1}\hbar)}\cr
&= \sum_{m\in\Z} e^{\cP_m} \sz^m \left\{ \sz \left(\frac{\hbar}{q}\right)^{\frac{L}{2}} \prod_{i=1}^L \frac{1-q^{-m}s^{-1}x_i^{-1}}{1-q^{-m-1}s^{-1} x_i^{-1}\hbar} \right\} \prod_{i=1}^L \frac{\upphi(q^{1-m}s^{-1}x_i^{-1})}{\upphi(q^{-m}s^{-1} x_i^{-1}\hbar)}.
\end{align}
Triviality of the insertion in curly brackets implies the following difference equation:
\begin{equation}
\left(\prod_{i=1}^L (1-W^{-1}x_i^{-1})-
\sz_\#^{-1}\prod_{i=1}^L (1-W^{-1} x_i^{-1}\hbar)\right) \tilde{V}_p=0,
\end{equation}
where $\sz_\# = \sz (\hbar/q)^{\frac{L}{2}}$. We recognize the quantum equivariant K-theory relation of $T^*\C\bP^{L-1}$, with $\sz_\#^{-1}$ being the quantum parameter, so that $\sz_\#^{-1}\to 0$ is the classical limit. Such relations can be also read off the effective twisted superpotential \cite{Nekrasov:2009uh,Nekrasov:2009ui}.

\subsection{K-theory and elliptic cohomology}
\subsubsection{Vertex with descendants}
Recall that the Vertex function was defined as a $\C^\times_q$-equivariant integral over the moduli of quasimaps from $\C\bP^1$ to $X$ regular at $[\infty]\in\C\bP^1$. The Vertex $V\in K_{\bfT\times\C^\times_q}(X)_{\rm loc}\otimes \Q[[\sz]]$ was ``read off'' at $[\infty]$ via the evaluation map ${\rm ev}_\infty$. Physically, $V$ is computed by the 3d A-twisted background $S^1\times \C\bP^1$, with some condition at $[\infty]\in\C\bP^1$ (and abelian Wilson loops inserted for convenience). We can remove a neighborhood of $[\infty]$ (turning $\C\bP^1$ into cigar), and roughly read off $V$ as the output state at the boundary torus, $V\in\cH_A[\mathbb{T}^2]$. Alternatively, like earlier in this paper, we can decompose $V$ in a fixed point basis of $K_{\bfT\times\C^\times_q}(X)_{\rm loc}$. This amounts to finding the coefficients $V_p$ computed by the $S^1\times \C\bP^1$ partition function, with the fields along $S^1\times[\infty]\in S^1\times\C\bP^1$ constrained to take values corresponding to the fixed point $p$.

We also saw that an alternatively normalized Vertex function $\tilde{V}$ was more naturally interpreted as the answer provided by the half-index. One can still think of it as an element of K-theory, however, as it is evaluated against the $(0,2)$ boundary conditions, $\tilde{V}$ is more naturally viewed as taking values in the equivariant elliptic cohomology \cite{Dedushenko:2021mds,Bullimore:2021rnr}. Namely, the boundary anomaly determines a line bundle $\cL$ over the extended elliptic cohomology variety ${\rm E}_T(X)$, and $\tilde{V}$ gives its holomorphic section.

We also saw in Section \ref{sec:quasimaps} following \cite{Okounkov:2015spn} that there exists a natural generalization called vertex with descendant, corresponding to inserting $\lambda\in K_G({\rm pt})$ at $[0]\in\C\bP^1$. Equivalently, in \cite{Okounkov:2020nql} this object is defined by pushing and pulling via ${\rm ev}_\infty$ and ${\rm ev}_0$. A quasimap may be singular at $[0]$, so unlike ${\rm ev}_\infty$, the evaluation map ${\rm ev}_0$ at $[0]$ does not land in $X$. Rather, it maps to the quotient stack:
\begin{equation}
{\rm ev}_0: {\rm QM}_{{\rm nonsing},\infty}(X) \to \mathscr{X} = \left[ \mu_\C^{-1}(0)/G\right].
\end{equation}
One can therefore include insertions at $[0]$ by pulling back the K-theory classes of $\mathscr{X}$ by ${\rm ev}_0$. Note that
\begin{equation}
K(\mathscr{X})\cong K_G(\mu_\C^{-1}(0)) \cong K_G({\rm pt}),
\end{equation}
where in the last step we used that $\mu_\C^{-1}(0)$ is equivariantly contractible, at least for the complex moment map appearing in 3d $\cN=4$ theories. Thus the insertions of descendants $\lambda\in K_G({\rm pt})$ that we saw in Section \ref{sec:quasimaps} are the same as pulling back the K-theory classes from the quotient stack $\mathscr{X}$ via the evaluation map ${\rm ev}_0$. In fact, in \cite{Okounkov:2020nql} the Vertex with descendant is defined precisely in such a way, resulting in a map $K_{\bfT\times\C^\times_q}(X)\to K_G(\mu_\C^{-1}(0))_{\rm loc}[[\sz]]$ or $K_G(\mu_\C^{-1}(0))\to K_{\bfT\times\C^\times_q}(X)_{\rm loc}[[\sz]]$, depending in which direction we pull-push. We prefer to consider the insertion of a class from $K(\mathscr{X})$ as an ``input'', so we have:
\begin{equation}
({\rm ev}_\infty)_* \left( \sz^{\rm deg} \hat{\cO}_{\rm vir}\otimes {\rm ev}_0^*(\cdot) \right): K_G(\mu_\C^{-1}(0)) \to K_{\bfT\times\C^\times_q}(X)_{\rm loc}\otimes \Q[[\sz]].
\end{equation}

If we work in the normalization $\tilde{V}$, such Vertex also admits descendants. Being computed by the half-index with $(0,2)$ boundary, it maps K-theory classes to the elliptic cohomology classes valued in the line bundle $\cL$ determined by the boundary anomalies:
\begin{equation}
\label{halfIndDesc}
\tilde{V} \text{ with descendant}: K_G(\mu_\C^{-1}(0)) \to H^0({\rm E}_T(X), \cL).
\end{equation}
We could also include $\bfT$-equivariance in the domain of this map, if necessary, namely, consider $K_{G\times \bfT}(\mu_\C^{-1}(0))$.

\subsubsection{Role of lines and surfaces}
Since the descendants were also argued in Section \ref{sec:quasimaps} to correspond, physically, to the insertions of supersymmetric Wilson loops at $[0]$, we see that such Wilson loops represent the elements of $K(\mathscr{X})$ or $K_\bfT(\mathscr{X})$. Thus the physical content of \eqref{halfIndDesc} is simply the half-index, with the Wilson loop representing $\lambda\in K(\mathscr{X})$ placed at the tip (the ``input''), and the state representing some elliptic cohomology class read off at the boundary (the ``output'').

Wilson loops labeled by $\lambda\in K_G({\rm pt})=K(\mathscr{X})$ satisfy the relations, like those derived in the examples of previous subsection, which carve out the quantum K-theory ring of $X$ from $K(\mathscr{X})$. Thus we can write:
\begin{equation}
QK(X) = K(\mathscr{X})/(\text{relations on the Wilson loops}).
\end{equation}


As for the role of surface operators $\Sigma$ introduced earlier, we saw that they act on the vertex function, modifying each $\tilde{V}_p$ by a certain elliptic class written as a ratio of theta-functions in \eqref{Sigma_action}. That elliptic class, though an elliptic function of $s$, is in general a non-trivial section in its dependence on the global connections $(\hbar,x,z)$. Thus it maps elliptic classes in $H^0({\rm E}_T(X),\cL)$ to $H^0({\rm E}_T(X),\cL')$, where ${\cL'} \otimes {\cL}^{-1}$ is determined by (the factors of automorphy of) $\Sigma$. We could be even more general and relax all anomaly-cancellation conditions imposed on $\Sigma$. Such more general $\Sigma$ thus cannot be inserted in the bulk of $S^1\times HS^2$, only along its boundary, and only if we have Dirichlet boundary conditions on the gauge fields (so that there is no possible gauge anomaly along the boundary). Thus the most general setup can be summarized by the following sequence of maps:
\begin{equation}
K_G(\mu_\C^{-1}(0)) \xrightarrow{\tilde{V}} H^0({\rm E}_T(X), \cL) \xrightarrow{\Sigma} H^0({\rm E}_T(X), \cL').
\end{equation}
In this context, we can think of $\Sigma$ as realized by boundary degrees of freedom that modify the boundary conditions, e.g., Fermi multiplets ``flipping'' the boundary conditions in \cite{Dimofte:2017tpi}.

\section{Outlook}
The cigar partition function and its close relatives, such as the half-index, hemisphere partition function, and their twisted versions, sit at the crossroads of many subfield of mathematical physics. In this work we have highlighted its roles in the vortex or quasimap counting, and in probing the 3d mirror symmetry and symplectic duality. Our results identify the precise physical home for several mathematical constructions, and also use physics to point towards new objects. Among other things, we also saw that the cigar supersymmetric index connects elliptic cohomology with the quantum K-theory.

Physically, quantum K-theory comes about as the algebra of supersymmetric Wilson loops in 3d \cite{Kapustin:2013hpk}, generalizing quantum cohomology in 2d \cite{Witten:1993xi}. Such algebras can often be read off from the effective twisted superpotential as in \cite{Nekrasov:2009uh,Nekrasov:2009ui,Nekrasov:2009rc}. Their relations also manifest themselves as the difference equations on the half-index, or K-theoretic J-function, as exhibited in \cite{Jockers:2018sfl} and in a couple of examples above. See also references \cite{Gu:2020zpg,Gu:2022yvj}. These developments, on the one hand, seem to connect with the mathematical approach of \cite{Givental:2017fng}. One the other hand, the interpretation of half-index in terms of the quasimap-counting Vertex function suggests connection to the quasimap-based approach to the quantum K-theory \cite{Okounkov:2015spn,Koroteev:2017nab,Pushkar:2016qvw}.

The precise relation between these approaches, though perhaps known to some experts, is not spelled out yet. The physical aspects are especially interesting here. In the quasimap approach, the quantum product is defined using the so-called relative insertions. These insertions allow to impose regularity on quasimaps at a point $p\in C$ without breaking the compactness of ${\rm QM}(X)$, and without breaking the properness of the evaluation map ${\rm ev}_p$. Thus such insertions, without the need to resort to the $\C^\times_q$-equivariance, allow to pull-push classes from $K(X)$ and use them to define the quantum product.

Notice that we have not discussed the physical meaning of relative insertions in this paper, because we do not have one at the moment. It is also known that such insertions are connected to the descendant insertions that we did discuss earlier, and the relation goes via the K-theoretic stable envelopes \cite{Aganagic:2017gsx}. The mathematical construction of relative insertions \cite{Okounkov:2015spn} involves gauging the $\C^\times_q$ automorphism of $\C\bP^1$, which suggests that it might involve activating fields of the supergravity background. It would be interesting to understand the physical meaning of relative insertions.

Besides that, the current work, together with our previous paper \cite{Dedushenko:2021mds}, form stepping stones towards the part III of this series. It will be devoted to clarifying the meaning of Bethe/Gauge correspondence, the role of interfaces in it, and its String-theoretic origins. It also appears important to make contact with the following sequence of papers: \cite{Galakhov:2020vyb,Galakhov:2021xum,Galakhov:2021vbo,Galakhov:2022uyu,Li:2023zub,Galakhov:2023aev}.

\appendix

\section{A-twisted background}\label{app:Atwist}
By taking the limit \eqref{GL_stretch} of the squashed background in Section \ref{sec:stretch}, we obtain the A-twisted theory on the cigar $S^1\times_\varepsilon C$. It can also be written using the standard methods, as in \cite{Benini:2015noa,Benini:2016hjo}, and the results agree. As we know, the background has $A^{(R)}_\varphi = \frac12 \omega_\varphi$, which we keep as a background field, i.e., in this Appendix we do not rewrite the theory in twisted variables. The effect of the limit is to simply drop various $\sim\frac1{f(\theta)}$ terms in the SUSY and actions. The SUSY variations of the vector multiplet are:
\begin{align}
\delta A_\mu &= \frac{i}{2} (\bar\epsilon \gamma_\mu\lambda + \epsilon\gamma_\mu \bar\lambda)\cr
\delta \sigma &= \frac12 (\bar\epsilon\lambda - \epsilon\bar\lambda)\cr
\delta\lambda&= -\frac12 F_{\mu\nu} \gamma^{\mu\nu}\epsilon - D\epsilon + i\partial_\mu\sigma \gamma^\mu\epsilon\cr
\delta\bar\lambda&= -\frac12 F_{\mu\nu}\gamma^{\mu\nu}\bar\epsilon + D \bar\epsilon -i\partial_\mu \sigma \gamma^\mu \bar\epsilon\cr
\delta D&= -\frac{i}{2}\bar\epsilon \gamma^\mu D_\mu\lambda + \frac{i}{2} \epsilon \gamma^\mu D_\mu\bar\lambda + \frac{i}{2}\bar\epsilon [\lambda,\sigma] + \frac{i}{2}\epsilon[\bar\lambda,\sigma].\cr
\end{align}
SUSY of the chiral multiplet:
\begin{align}
\delta \phi &=\bar\epsilon \psi,\quad \delta \bar\phi=\epsilon \bar\psi\cr
\delta\psi&=i \gamma^\mu \epsilon D_\mu\phi + i\epsilon\sigma\phi + \bar{\epsilon} F\cr
\delta\bar\psi&=i\gamma^\mu \bar\epsilon D_\mu\bar\phi + i\bar\phi \sigma\bar\epsilon + \bar{F}\epsilon\cr
\delta F&= \epsilon(i\gamma^\mu D_\mu\psi -i\sigma\psi -i\lambda\phi)\cr
\delta \bar{F}&=\bar\epsilon(i\gamma^\mu D_\mu\bar\psi-i\bar\psi\sigma+i\bar\phi\bar\lambda).
\end{align}
Invariant kinetic actions:
\begin{align}
\cL_{\rm v}=\Tr\Big[ \frac12 F^{\mu\nu}F_{\mu\nu} + D^\mu\sigma D_\mu\sigma+ D^2 -\epsilon^{\mu\nu\rho}F_{\mu\nu}D_\rho\sigma\Big]+\Tr\Big[ i\lambda\gamma^\mu D_\mu\bar\lambda + i\lambda[\bar\lambda,\sigma] \Big].
\end{align}
\begin{align}
\cL_{\rm c}=-\bar\phi D^\mu D_\mu\phi +\bar\phi \sigma^2 \phi + i\bar\phi D\phi - \bar\phi (\Delta \star_2\dd A^{(R)})\phi + \bar{F} F -i\bar\psi\gamma^\mu D_\mu \psi + i\bar\psi\sigma\psi  +i\bar\psi\lambda\phi -i\bar\phi\bar\lambda\psi,
\end{align}
where as before, $\Delta$ denotes the R-charge, and covariant derivatives include the background R-symmetry gauge field $A^{(R)}$. The superpotential action is the same as in flat space.

\section{Localizing deformations on $S^1\times S^2_b$}\label{app:LocDef}
Here we review a few standard deformations that can be used to localize the $S^1\times S^2_b$ index. The canonical localizing term for the vector multipelts is defined by
\begin{equation}
L^v_{\rm Loc} = \frac12 \sum_{\alpha} \delta\left[ \lambda_\alpha (\delta\lambda_\alpha)^* + \bar\lambda_\alpha (\delta\bar\lambda_\alpha)^* \right] = L^v_{\rm Bos} + L^v_{\rm Ferm},
\end{equation}
where the bosonic part can be written as
\begin{align}
L^v_{\rm Bos}&=D^2 + D_\mu\sigma D^\mu \sigma + \frac{(\ell\sigma\sin\theta-F_{\theta\varphi})^2}{f(\theta)^2\ell^2\sin^2\theta} + \frac{(F_{\varphi\alpha})^2}{\beta^2\ell^2\sin^2\theta} + \frac{(F_{\theta\alpha} - \varepsilon F_{\theta\varphi})^2}{\beta^2f(\theta)^2}.
\end{align}
For the chiral multiplets, we similarly define
\begin{equation}
\label{canPsi}
L^\chi_{\rm Loc} = \frac12 \sum_{\alpha} \delta\left[ \psi_\alpha (\delta\psi_\alpha)^* + \bar\psi_\alpha (\delta\bar\psi_\alpha)^* \right] = L^\chi_{\rm Bos} + L^\chi_{\rm Ferm},
\end{equation}
and the bosonic part is:
\begin{align}
L^\chi_{\rm Bos} &= F\bar{F} + \frac1{\beta^2}|D_\alpha\phi - \varepsilon D_\varphi\phi|^2 + \left| \frac{i}{\ell}D_\varphi\phi - \frac{\Delta}{f(\theta)}\phi -\hat\sigma\phi\cos\theta \right|^2\cr 
&+ \left| \frac1{f(\theta)} D_\theta\phi +\hat\sigma\phi\sin\theta + \frac{i}{\ell}\cot\theta D_\varphi\phi \right|^2.
\end{align}
Here
\begin{equation}
\hat\sigma=\sigma + m
\end{equation}
includes both the gauge multiplet scalar $\sigma$ and the flavor mass $m$. In taking the absolute values squared, we assume the reality conditions $\bar\phi = \phi^*$, $\sigma^*=\sigma$ and $m^*=m$. In deriving this action from \eqref{canPsi}, we also assumed that $\bar{F} = F^*$, though this is not necessary for the Q-exactness of the final expression. The same comment applies to the vector part $L^v_{\rm Bos}$, where $D$ does not have to be real.

One can also define the following term \cite{Benini:2013yva,Fujitsuka:2013fga}:
\begin{align}
L_{H}&= i\cQ \Big[(\bar\epsilon\gamma_{\hat3}\lambda - \epsilon\gamma_{\hat3}\bar\lambda)H(\phi) \Big]\cr 
&=-2H(\phi) 
\left( iD + \frac{1}{f(\theta)}D_\theta(\sigma\sin\theta) - \frac{\cot\theta}{\ell f(\theta)}F_{\theta\varphi}\right) + {\rm ferm},
\end{align}
where it is customary to take $H(\phi)$ as the real moment map:
\begin{equation}
H(\phi)= e^2\mu_\R=e^2(\bar\phi \phi - \zeta).
\end{equation}
The modified localizing term for gauge multiplets:
\begin{align}
L^v_{\rm Bos}+L_{H}\big|_{\rm bos}&=(D-iH(\phi))^2 + \frac1{\ell^2\sin^2\theta}D_\varphi\sigma D_\varphi\sigma + \frac1{\beta^2}(D_\alpha\sigma - \varepsilon D_\varphi\sigma)^2\cr 
&+ \left(H + \frac1{f(\theta)}\left( \frac{F_{\theta\varphi}\cos\theta}{\ell\sin\theta} - D_\theta(\sigma\sin\theta) \right) \right)^2 + \frac1{f(\theta)^2}\left(D_\theta(\sigma\cos\theta) + \frac{F_{\theta\varphi}}{\ell}  \right)^2\cr & + \frac{(F_{\varphi\alpha})^2}{\beta^2\ell^2\sin^2\theta} + \frac{(F_{\theta\alpha} - \varepsilon F_{\theta\varphi})^2}{\beta^2 f(\theta)^2}.
\end{align}
The deformation $L^v$ alone is used for the ``Coulomb branch localization'', while $L^v + L_{H1}$ leads to the ``Higgs branch localization''. The parameters $e$ and $\zeta$ are like the gauge coupling and the FI parameter, but more precisely, they are just parameters in the localizing deformation. For the Higgs branch localization, one has to take $\zeta\to\infty$ (note that the physical FI parameter also grows in the IR).

We can also alter the matter localization pattern by defining the following expression:
\begin{equation}
L_W=\cQ \left[ e^{i\nu} \partial_i W(\phi) \bar\epsilon.\psi^i - e^{-i\nu}\partial_i \overline{W}(\bar\phi) \epsilon.\bar\psi  \right],
\end{equation}
where we used the notation $\bar\epsilon.\psi = \sum_\alpha \bar\epsilon_\alpha \psi_\alpha$, and analogously for $\epsilon.\bar\psi$. This $L_W$ is not quite a superpotential term -- the superpotential (that is part of the physical action) is not $\cQ$-exact. However, it models the effect of the superpotential, and we find that:
\begin{align}
L_{\rm Bos}^\chi + L_W\big|_{\rm bos} &= (F-e^{-i\nu}\bar\partial\overline{W})(\bar{F}+e^{i\nu}\partial W) + \frac1{\beta^2}|D_\alpha\phi - \varepsilon D_\varphi\phi|^2 + \left| \frac{i}{\ell}D_\varphi\phi - \frac{\Delta}{f(\theta)}\phi -\hat\sigma\phi\cos\theta \right|^2\cr 
&+ \left| \frac1{f(\theta)} D_\theta\phi +\hat\sigma\phi\sin\theta + \frac{i}{\ell}\cot\theta D_\varphi\phi - ie^{-i\nu}\bar\partial\overline{W} \right|^2.
\end{align}
When using this term for localization, we can first integrate out $F, \bar{F}$, and end up with a positive definite action again, which now depends on $W$.

\paragraph{Coulomb branch localization} is achieved using $L^v$ and $L^\chi$ as the deformations. The localization equations for the vector multiplets are:
\begin{align}
D&=0,\quad D_\mu\sigma=0,\quad F_{\varphi\alpha}=0,\cr
F_{\theta\alpha}&=\varepsilon F_{\theta\varphi},\quad F_{\theta\varphi}=\ell\sigma \sin\theta,\cr
\end{align}
and the chirals obey:
\begin{align}
\label{MatterLoc}
F &= \bar{F} =0,\quad D_\alpha\phi =\varepsilon D_\varphi\phi,\cr
0&= \frac{i}{\ell}D_\varphi\phi - \frac{\Delta}{f(\theta)}\phi -\hat\sigma\phi\cos\theta,\cr
0&= \frac1{f(\theta)} D_\theta\phi +\hat\sigma\phi\sin\theta + \frac{i}{\ell}\cot\theta D_\varphi\phi.
\end{align}

\paragraph{Higgs branch localization} is obtained by replacing $L^v$ by $L^v + L_H$, which modifies the equations for vector multiplets: 
\begin{align}
\label{HBL1}
D_\varphi\sigma &= D_\alpha\sigma =0,\quad \ell D_\theta(\sigma\cos\theta) + F_{\theta\varphi}=0,\cr
F_{\varphi\alpha}&=0,\quad F_{\theta\alpha} - \varepsilon F_{\theta\varphi}=0,\cr
H(\phi)&= \frac{D_\theta(\sigma\sin\theta)}{f(\theta)} - \frac{F_{\theta\varphi}\cos\theta}{\ell f(\theta)\sin\theta},
\end{align}

\setlength{\unitlength}{1mm}

\newpage

\bibliographystyle{utphys}
\bibliography{refs}
\end{document}